\documentclass[11pt]{article}
\pdfoutput=1 
\usepackage{microtype}
\usepackage{mathtools}
\usepackage{booktabs}
\usepackage[english]{babel}
\usepackage{amsmath,amssymb,amsbsy,amstext, amsthm, simplewick, amsfonts}
\usepackage{graphicx}
\usepackage[small]{caption}
\usepackage{siunitx}
\usepackage{upgreek}
\usepackage{framed}
\usepackage{wrapfig}
\usepackage{multirow}
\usepackage{bbm}
\usepackage[numbers,sort&compress]{natbib}
\usepackage[svgnames,dvipsnames,x11names]{xcolor}
\usepackage[utf8x]{inputenc}
\usepackage{selinput}
\usepackage{bm}
\usepackage{float}
\usepackage[top=2.5cm, bottom=2.5cm, left=3cm, right=3cm]{geometry}
\usepackage{yfonts}
\usepackage{caption}
\usepackage{subcaption}
\usepackage{sidecap}
\usepackage{longtable}
\usepackage{anyfontsize}
\usepackage[hypertexnames = false]{hyperref}
\usepackage{cleveref}

\usepackage{epstopdf}
\usepackage{cancel}
\usepackage{tcolorbox}

\usepackage{pgfornament}

\newcommand{\sdg}{\sqrt{-g}}
\newcommand{\mn}{{\mu\nu}}

\newcommand{\Ij}{{ij}}

\newcommand{\rp}{r_+}
\newcommand{\rmn}{r_-}
\newcommand{\dd}{\mathrm{d}}
\newcommand{\xx}{x}
\newcommand{\rr}{r^2}

\DeclareFontFamily{OT1}{rsfs10}{}
\DeclareFontShape{OT1}{rsfs10}{m}{n}{ <-> rsfs10 }{}
\DeclareMathAlphabet{\mathscript}{OT1}{rsfs10}{m}{n}

\newcommand{\be}{\begin{equation}}
\newcommand{\ee}{\end{equation}}
\newcommand{\bea}{\begin{eqnarray}}
\newcommand{\eea}{\end{eqnarray}}
\newcommand{\bmat}{\begin{bmatrix}}
\newcommand{\emat}{\end{bmatrix}}
\newcommand{\beq}{\begin{equation}}
\newcommand{\eeq}{\end{equation}}
\newcommand{\beqa}{\begin{eqnarray}}
\newcommand{\eeqa}{\end{eqnarray}}
\newcommand{\beqar}{\begin{eqnarray*}}
\newcommand{\eeqar}{\end{eqnarray*}}

\newcommand{\bbibitem}[1]{\bibitem{#1}\marginpar{#1}}

\def\Label#1{\label{#1}%
  \smash{\hbox to0pt{\raise1ex\hbox{\tiny[#1]}\hss}}}
\def\noLabels{\let\Label=\label}
\def\nobbibitem{\let\bbibitem=\bibitem}

\newcommand{\e}{{\rm e}}
\newcommand{\D}{{\rm d}}

\newcommand{\fra}{\mathfrak{a}}
\newcommand{\frb}{\mathfrak{b}}
\newcommand{\frc}{\mathfrak{c}}
\newcommand{\hl}{{\hat\ell}}

\newcommand{\hypergeom}[2]{
  \mathbin{_{#1}{\sf F}_{#2}} }

\makeatletter
\DeclareRobustCommand{\rcite}[1]{%
  \rcite@aux#1,\@nil{#1}%
}
\def\rcite@aux#1,#2\@nil#3{%
  \if\relax#2\relax
    Ref.~\cite{#3}%
  \else
    Refs.~\cite{#3}%
  \fi
}
\makeatother

\definecolor{greyish}{rgb}{.90,.90,.90}
\definecolor{greyish2}{rgb}{.96,.96,.96}
\usepackage{xcolor,colortbl}
\usepackage{tcolorbox}

\numberwithin{equation}{section}

\begin{document}
%
\renewcommand{\thefootnote}{\fnsymbol{footnote}}
\vspace{0truecm}
\thispagestyle{empty}

\hfill

\begin{center}
{\fontsize{21}{18} \bf Love Numbers for Rotating Black Holes}\\[14pt]
{\fontsize{21}{18} \bf  in Higher Dimensions}
\end{center}

\vspace{.15truecm}

\begin{center}
{\fontsize{13}{18}\selectfont
Maria J.~Rodriguez,$^{\rm a,b}$\footnote{\texttt{majo.rodriguez.b@gmail.com}} Luca Santoni,$^{\rm c}$\footnote{\texttt{santoni@apc.in2p3.fr}} }\\[4.5pt]
{\fontsize{13}{18}\selectfont
Adam R.~Solomon,$^{\rm d,e}$\footnote{\texttt{soloma2@mcmaster.ca}} and 
Luis Fernando Temoche$^{\rm a}$\footnote{\texttt{l.f.temoche@usu.edu}}
}
\end{center}
\vspace{.4truecm}

 \centerline{{\it ${}^{\rm a}$Department of Physics, Utah State University,}}
 \centerline{{\it 4415 Old Main Hill Road, UT 84322, U.S.A.}}

 \vspace{.3cm}

 \centerline{{\it ${}^{\rm b}$Instituto de Fisica Teorica UAM/CSIC, Universidad Autonoma de Madrid,}}
 \centerline{{\it 13-15 Calle Nicolas Cabrera, 28049 Madrid, Spain}}

 \vspace{.3cm} 

 \centerline{{\it ${}^{\rm c}$Universit\'e Paris Cit\'e, CNRS, Astroparticule et Cosmologie,}}
 \centerline{{\it 10 Rue Alice Domon et L\'eonie Duquet, F-75013 Paris, France}} 
 
  \vspace{.3cm}
 
 \centerline{{\it ${}^{\rm d}$Department of Physics and Astronomy, McMaster University, }}
 \centerline{{\it 1280 Main Street West, Hamilton ON, Canada}}
 
 \vspace{.3cm}
 
 \centerline{{\it ${}^{\rm e}$Perimeter Institute for Theoretical Physics,}}
 \centerline{{\it 31 Caroline Street North, Waterloo ON, Canada}}

 \vspace{.25cm}

\vspace{.3cm}
\begin{abstract}
\noindent

We compute the tidal Love numbers and static response coefficients associated to several rotating black holes in higher dimensions, including Myers--Perry black holes, black rings, and black strings. These coefficients exhibit a rich and complex structure as a function of the black hole parameters and multipoles. Our results agree in limiting cases with known and new expressions for various lower-dimensional black holes. In particular, we provide an alternative approach to the computation of the static response of Kerr black holes as a limiting case of the boosted black string.

\end{abstract}

\newpage

\setcounter{tocdepth}{2}
\tableofcontents
\newpage
\renewcommand*{\thefootnote}{\arabic{footnote}}
\setcounter{footnote}{0}




\section{Introduction}
\label{Sec:Intro}

The tidal Love numbers are a set of quantities that characterize the conservative static response of a gravitating object under the influence of an external tidal field. As an intrinsic property of black holes and other compact objects, the Love numbers have been studied extensively in recent years due to the role they play in gravitational-wave astronomy: during the inspiral phase of a binary merger, the tidal coupling between the two compact objects can leave observable imprints on the waveform.

How an object deforms tidally is related to what it is made of, and indeed measurements of neutron star Love numbers are expected to provide new constraints on the equation of state \cite{Flanagan:2007ix,Raithel:2018ncd,Chatziioannou:2020pqz}. For black holes in four-dimensional general relativity, however, the Love numbers do not appear to say very much about internal structure, because they vanish regardless of the black hole's mass and spin, for both gravitational-wave polarizations \cite{Damour:2009va,Damour:2009vw,Binnington:2009bb,Kol:2011vg,Porto:2016zng,Poisson:2020mdi,Hui:2020xxx}. The static response coefficients turn out in fact to be purely imaginary, corresponding to dissipative effects induced by the rotation of the black hole \cite{LeTiec:2020spy, LeTiec:2020bos,Chia:2020yla,Charalambous:2021mea}.  The vanishing of the real part, i.e., of the Love numbers, in four dimensions instead reveals underlying hidden symmetries of general relativity \cite{Hui:2021vcv,Hui:2022vbh,Charalambous:2021kcz,Charalambous:2022rre}, indicating their potential as a tool to better understand gravitational dynamics.

In dimensions greater than four, the structure of the Love numbers becomes more intricate, vanishing for specific multipoles, while other multipoles can display features such as running \cite{Kol:2011vg,Hui:2020xxx}. This complexity reflects the rich geometry of higher-dimensional spacetimes and highlights the need for a deeper understanding of black hole physics in these contexts. Ultimately, the study of Love numbers in higher-dimensional black holes can shed light on the fundamental nature of gravity and the behavior of gravitational waves.

In this paper, we aim to further map out the behavior of the induced static response and  Love numbers in the zoo of higher-dimensional rotating black holes.\footnote{See also \rcite{Chakravarti:2018vlt,Charalambous:2023jgq}.} We consider three classes of solutions, distinguished by the topology of their horizons: rotating Myers--Perry black holes in $D=5$, black rings, and boosted black strings. In these cases, in suitable regimes the relevant equation of motion can be put into hypergeometric form, allowing for explicit solutions from which the Love numbers can be read off.

Strictly speaking, to determine the Love numbers we should solve the Einstein equations linearized about a black hole background, i.e., the equations of a massless spin-2 field on said background. It turns out that the qualitative features of the Love numbers, such as their multipolar and dimensional dependence, are often largely independent of the field's spin, and so we can simplify matters significantly by working instead with the Klein--Gordon equation for a massless scalar \cite{Kol:2011vg,Hui:2020xxx,Hui:2021vcv,Charalambous:2021mea}.\footnote{There are however interesting exceptions, see, e.g., \rcite{Hui:2020xxx,Pereniguez:2021xcj}.} Therefore we should emphasize that in this paper we are calculating \emph{scalar} Love numbers.

This paper provides an in-depth analysis of the intersection of tidal deformations and higher-dimensional black holes in vacuum general relativity (GR). \Cref{Sec:methods} gives an overview of the methods we use in each of the spacetime backgrounds we consider. \Cref{Sec:MP5D} discusses these coefficients for Myers--Perry black holes in five spacetime dimensions. In \cref{Sec:BR} we discuss the static Love numbers in a black ring background, including separability of the Klein--Gordon equation, new problem-solving approaches, and limiting cases. The static Love numbers for black strings in near-zone approximation are explored in \cref{Sec:BS}. Finally, the discussion in \cref{Sec:Discussion} provides a summary of the key takeaways for the static Love numbers in higher-dimensional GR and speculations on the point-particle effective field theory approach. 

{\it Notation and conventions:} We work in natural units $c=\hbar = 1$ (though leave in $G$) and use the mostly positive signature for the metric. In the text, we use the same symbol $R$ to denote two different quantities: the radial component of the scalar field in \cref{Sec:MP5D,Sec:BS} (see, e.g., \cref{Psi5DKG}), and one of the parameters of the black ring metric in \cref{Sec:BR} (see \cref{neutral}). In addition, the same symbol $\Delta$ is used in the expressions of the various metrics that we consider in this paper: the reader should refer to the definition that is given within each section for  $\Delta$.


\section{Methods}
\label{Sec:methods}

As a proxy for linear tidal responses in gravity, this paper considers static solutions to the Klein--Gordon (KG) equation for a massless scalar $\Psi(x)$ living on a $D$-dimensional stationary spacetime background $g_\mn$,
\be\label{KGini}
\Box \Psi(x) = \frac{1}{\sdg} \partial_\mu \left[ \sqrt{-g}g^{\mu\nu}\partial_\nu\Psi(x)\right]=0.
\ee
We will mostly be interested in solutions that have Cauchy horizons, such as black holes and black rings. To begin with, let us assume that the background has $D-2$ Killing vectors, one timelike and the rest spacelike. We can therefore choose a coordinate basis $x^\mu=(t,r,\theta,\phi_k)$, where the index $k$ runs over $1,\cdots,D-3$, so the Killing vectors correspond to $\partial_t, \partial_{\phi_k}$. The invariance of \cref{KGini} under the isometries generated by these Killing vectors allows us to decompose the field as 
\be\label{fullWV}
\Psi(x)= \exp\left(-i\omega t + i \displaystyle\sum^{D-3}_{k=1}   m_{k}\phi_k\right) \Phi(r,\theta)\,.
\ee
We will be interested in cases where the solution fully separates,
\begin{equation}
\Phi(r,\theta)=R(r)Y(\theta).
\end{equation}
This separability property famously holds for Kerr black holes in $D=4$ and persists to higher-dimensional (Myers--Perry) black holes \cite{Frolov:2006pe}. In both cases this is due to a ``hidden" symmetry generated by one or more Killing tensors \cite{Frolov:2017kze}. The situation is more subtle for backgrounds with more general horizon topologies, such as black rings and black strings. However, it turns out that for the static field configurations ($\omega=0$) that we are interested in, the KG equation does separate in these backgrounds \cite{Cardoso:2005sj}. The object of our study will therefore be the radial equation of motion for $R(r)$ in a variety of higher-dimensional black hole backgrounds.

The radial equation is a second-order ordinary differential equation, so has two independent solutions. The physical situation we have in mind is a black hole immersed in an external, static tidal field.
At infinity, if the metric is asymptotically flat, the solutions to the radial equation can either grow as $r^\ell$ or decay as $r^{-\ell-n}$, where we have assumed that the solutions $Y(\theta)=Y_\ell(\theta)$ are spherical harmonics on an $n+1$-sphere.\footnote{We have left $n$ general here since its relationship to $D$ will depend on the horizon topology. For instance, $n=D-3$ for a black hole, but $n=D-4$ for a black string in the near region (see \cref{Sec:BS}).} 
The growing behavior would not be physical if infinity were truly infinity, but here we are taking it to be a proxy for the location of the tidal source. Decomposing the external tidal field in terms of a superposition of $r^\ell$ modes sets one boundary condition for each mode.

Meanwhile, at the (outer) horizon, one solution for $R(r)$ typically diverges logarithmically, while another can be chosen to approach a constant. For black holes, where the horizon is a physical location, we must discard the former solution on physical grounds.\footnote{To be more precise, one should require any diffeomorphism-invariant physical quantity built from the solution for $\Psi$ to be well-defined at the event horizon.} This fixes the other boundary condition. At infinity, this physical solution will be an admixture of growing and decaying modes,
\begin{equation}
R_\ell(r) \to R_{\ell,\infty}\left(r^\ell+\lambda_\ell r^{-\ell-n}\right),
\end{equation}
where $R_{\ell,\infty}$ a constant. The coefficient $\lambda_\ell$ is interpreted as the static tidal response coefficient.\footnote{Both the growing and decaying terms are the leading pieces in an expansion in $1/r$. This leads to an ambiguity in the definition of $\lambda_\ell$ when $\ell$ is an integer, as the $\mathcal{O}(r^{-2\ell-n})$ correction to the growing mode has the same $r^{-\ell-n}$ scaling as the leading static response. As we discuss in some more detail in \cref{Sec:BR}, this can be resolved by analytically continuing $\ell\in\mathbb{R}$, where the source-response split is unambiguous, extracting $\lambda_\ell$, and then taking $\ell\to\mathbb{N}$.} The Love number is typically defined to be the conservative part of the response, which is obtained by taking the real part of $\lambda_\ell$, while the imaginary piece corresponds to dissipative effects.


\section{$5D$ Myers--Perry Black Hole}
\label{Sec:MP5D}

We now turn to the Klein--Gordon wave equation \eqref{KGini} for the five-dimensional Myers--Perry black hole \cite{Myers:1986un}. The equations of motion in this geometry (in arbitrary dimensions) are separable \cite{Cvetic:1997uw,Krishnan:2010pv,Lu:2008jk} due to ``hidden symmetries" generated by a tower of Killing tensors \cite{Frolov:2006pe,Kubiznak:2006kt,Frolov:2006dqt,Frolov:2017kze}.\footnote{These symmetries are ``hidden" in the sense that they act on the full phase space of the dynamics, rather than the configuration space (spacetime). This results in conserved quantities which are non-linear in the momenta, in contrast to the ``explicit" symmetries generated by Killing vectors \cite{Frolov:2017kze}.} Since the aim is to compute Love numbers, which are static responses, we will consider static scalar field configurations.

\subsection{Background}

In many respects, the Myers--Perry solution describing spinning black holes in $D=5$ possesses the same remarkable properties as the standard Kerr black hole in four dimensions. They are the unique asymptotically-flat vacuum solutions with spherical topology, parametrized by their mass and two angular momenta. The metric in Boyer--Lindquist coordinates is
\begin{align}\label{eq:MPmetric}
\dd s^2 &= -\dd t^2+\frac{\mu}{\Sigma}\left(\dd t-a \sin^2\theta\,\dd\phi-b\cos^2\theta\,\dd\psi\right)^2 + \frac{r^2\Sigma}{\Delta}\dd r^2 + \Sigma \dd\theta^2 \nonumber\\
&\hphantom{{}=}+(r^2+a^2)\sin^2\theta \dd\phi^2 +(r^2+b^2)\cos^2\theta\dd\psi^2,
\end{align}
where
\begin{align}
\Sigma &= r^2+a^2\cos^2\theta+b^2\sin^2\theta,\\
\Delta &= (r^2+a^2)(r^2+b^2)-\mu r^2.
\end{align}
The coordinate ranges are $0<r\le \infty$, $0<\theta<\pi$ and $0\le \psi,\phi \le 2\pi$. These coordinates generalize the Boyer--Lindquist coordinates $(t,r,\theta,\phi)$ in $D=4$ by the addition of a second angular Killing direction $\psi$. There are three free parameters: $\mu$ is a mass parameter and $a$ and $b$ are rotation parameters, related to the physical mass $M$ and angular momenta $J_\phi$ and $J_\psi$ by
\begin{equation}
\mu = \frac{8GM}{3\pi}, \qquad J_\phi = \frac{2M}{3}a, \qquad J_\psi = \frac{2M}{3}b.
\end{equation}
There are two horizons, located at the roots of $\Delta=(r^2-r_+^2)(r^2-r_-^2)$,
\begin{equation}
2r_\pm^2 = \mu-a^2-b^2\pm\sqrt{(\mu-a^2-b^2)^2-4a^2b^2}.
\end{equation}
The existence of the horizons requires
\begin{align}
\mu &\geq a^2+b^2+2|ab|\\
\Longrightarrow M^3 &\geq \frac{27\pi}{32G}\left(J_\phi^2+J_\psi^2+2\left |J_\phi J_\psi \right|\right).
\end{align}

\subsection{Klein--Gordon equation}

To calculate the Klein--Gordon equation \eqref{KGini} for a static field $\Psi(r,\theta,\phi,\psi)$ in this geometry, it is helpful to retain manifest covariance on the 3-sphere, since we are ultimately most interested in the radial dynamics. To this end, let us package the coordinates on $S^3$ into $\theta^i=(\theta,\phi,\psi)$ with $i\in(1,2,3)$, and define the unit-sphere metric $\gamma_\Ij$ by $\dd\Omega^2=\gamma_\Ij\dd\theta^i\dd\theta^j=\dd\theta^2+\sin^2\theta\dd\phi^2+\cos^2\theta\dd\psi^2$. The KG equation requires the $5D$ metric determinant and the metric inverse in the angular directions,
\begin{equation}
\sdg = r\Sigma\sqrt\gamma,\qquad \Sigma g^\Ij = \gamma^\Ij +M^{ij}(r) ,
\end{equation}
where $\gamma^\Ij$ is the inverse of $\gamma_\Ij$, and the matrix $M^{ij}(r)$ is only non-zero along the Killing directions $(\phi,\psi)$:
\begin{equation}
M^\Ij\partial_i\partial_j=\frac{(b^2-a^2)(b^2+r^2)-b^2\mu}{\Delta}\partial_\phi^2+\frac{(a^2-b^2)(a^2+r^2)-a^2\mu}{\Delta}\partial_\psi^2-2\frac{ab\mu}{\Delta}\partial_\phi\partial_\psi.
\end{equation}
With these simplifications it is straightforward to check that the Klein--Gordon equation reduces (after multiplying by $\Sigma$) to
\begin{equation}
\frac1r\partial_r\left(\frac\Delta r\partial_r\right)\Psi + M^\Ij\partial_i\partial_j\Psi+ \nabla^2_{S^3}\Psi=0, \label{eq:KG1}
\end{equation}
where $\nabla^2_{S^3}$ is the Laplacian on the 3-sphere.

This equation is fully separable, admitting solutions of the form
\begin{align}
\Psi(x) &= \Theta(\theta^i)R(r) \\
&=\e^{i(m_\phi\phi+m_\psi\psi)}Y(\theta)R(r).
\label{Psi5DKG}
\end{align}
After dividing out by $\Psi$, the first two terms in \cref{eq:KG1} depend only on $r$ and the last only on $\theta$. This implies that $\Theta(\theta^i)$ is an eigenfunction of $\nabla^2_{S^3}$, i.e., a hyperspherical harmonic.\footnote{This is the higher-dimensional analogue of the fact that static perturbations of Kerr are expanded in (spin-weighted) spherical rather than spheroidal harmonics.} The separation constant is well-known to be $-\ell(\ell+2)$ for integer $\ell\geq0$,\footnote{See, e.g., App. A.1 of \rcite{Hui:2020xxx} and references therein.}
\begin{equation}
\nabla^2_{S^3}\Theta(\theta^i) = -\ell(\ell+2)\Theta(\theta^i),
\end{equation}
or equivalently,
\begin{equation}
\sec\theta\csc\theta\partial_\theta\left(\sin\theta\cos\theta\partial_\theta Y\right)-(m_\phi^2\csc^2\theta+m_\psi^2\sec^2\theta)Y = -\ell(\ell+2) Y.
\label{eq:Ytheta5D}
\end{equation}
With these simplifications we are only left with radial derivatives in \cref{eq:KG1}, so we can divide out $\Theta(\theta^i)$ to obtain the radial equation,
\begin{equation}
\boxed{\frac1r\partial_r\left(\frac\Delta r\partial_r\right)R - \left(\ell(\ell+2) + M^\Ij m_i m_j\right)R = 0,} \label{eq:KG}
\end{equation}
where $m_i\dd \theta^i=m_\phi\dd\phi+m_\psi\dd\psi$. We can write this more explicitly as
\begin{equation}
\frac \Delta r \partial_r\left(\frac \Delta r \partial_rR\right)+\left[(a m_\psi+b m_\phi)^2\mu +(a^2-b^2)(m_\phi^2(b^2+r^2)-m_\psi^2(a^2+r^2))\right]R=\ell(\ell+2)\Delta R.
\end{equation}
Herein we will replace $\ell$ with
\begin{equation}
\hat\ell \equiv \frac{\ell}{D-3} = \frac\ell2.
\end{equation}
This will turn out to be convenient because $\hl$ plays a role analogous to $\ell$ in $D=4$ \cite{Kol:2011vg,Hui:2020xxx}.

The radial equation \eqref{eq:KG} has five regular singular points, but two pairs of these (at the inner and outer horizons) are degenerate. By changing variables from $r$ to $r^2$ we can reduce the number of regular singular points to three, which guarantees that \cref{eq:KG} can be solved in terms of hypergeometric functions. Concretely, let us define the radial variable $x$ by
\begin{equation}
r^2 = \frac{(r_+^2-r_-^2)x+r_+^2+r_-^2}{2}.
\end{equation}
The inner and outer horizons are located at $x=-1$ and $x=+1$, respectively.
In terms of $x$ we have
\begin{equation}
\Delta = \frac14\left(r_+^2-r_-^2\right)^2(x^2-1),\quad \frac1r\partial_r = \frac{4}{r_+^2-r_-^2}\partial_x,
\end{equation}
so that the radial equation becomes
\begin{equation}
\partial_x\left[(x^2-1)\partial_xR\right] - \left[\hat\ell(\hat\ell+1) + \frac{1}{4} M^\Ij m_i m_j\right]R.\label{eq:KGx1}
\end{equation}

In order to solve the radial equation and read off the Love numbers, it is convenient to change the basis of $m_i$ to $(m_L,m_R)$ defined by
\begin{equation}
m_\phi = m_R + m_L,\quad m_\psi = m_R - m_L,
\end{equation}
and then to rescale each of these as\footnote{These prefactors can be expressed in terms of thermodynamic quantities: the angular velocities at the horizon $\Omega_{L,R}$ and the surface gravity at the outer horizon $\kappa_+$,
\[ \tilde{m}_{L,R} =\frac{\Omega_{L,R}}{\kappa_+} m_{L,R}.\]}
\begin{equation}
\tilde m_L = \frac{a-b}{\rp+\rmn}\frac{m_L}{2},\quad\tilde m_R = \frac{a+b}{\rp-\rmn}\frac{m_R}{2}.
\end{equation}
The term in \cref{eq:KGx1} involving $m_i$ factorizes into poles at the horizons $x=\pm1$,
\begin{equation}
-\frac{1}{4}M^\Ij m_i m_j = 2\left(\frac{(\tilde m_R + \tilde m_L)^2}{x-1} - \frac{(\tilde m_L- \tilde m_R)^2}{x+1}\right),
\end{equation}
so the Klein--Gordon equation takes the form
\begin{equation}
\boxed{\partial_x\left[(x^2-1)\partial_x R\right] + 2\left(\frac{(\tilde m_R + \tilde m_L)^2}{x-1} - \frac{(\tilde m_L- \tilde m_R)^2}{x+1}\right)R = \hat\ell(\hat\ell+1)R.} \label{eq:KGx}
\end{equation}

\subsection{Static responses}
\label{sec:5DMPstaticresponse}

The static Klein--Gordon equation \eqref{eq:KGx} has three regular singular points---the inner and outer horizons and infinity---and so admits hypergeometric solutions. The simplest solutions, which do not depend on the Killing directions, $m_\phi = m_\psi = 0$, are in fact Legendre polynomials,
\begin{equation}
R = c_1 P_{\hl}(x) + c_2 Q_{\hl}(x).
\end{equation}
To rewrite \cref{eq:KGx} in hypergeometric form we transform the radial variable,
\begin{equation}
z \equiv \frac{2}{1+x} = \frac{\rp^2-\rmn^2}{r^2-\rmn^2},
\end{equation}
and perform a field redefinition,
\begin{equation}
R(z) = z^{\hat\ell+1}(1-z)^{i(\tilde m_L+\tilde m_R)}u(z),
\label{eq:redRz}
\end{equation}
so that we obtain the standard hypergeometric equation,
\begin{equation}
z(1-z)u''(z)+\left[\mathfrak{c}-\left(\mathfrak{a}+\mathfrak{b}+1\right)z\right]u'(z)-\mathfrak{a\, b} \, u(z)=0
\label{eq:hstandard}
\end{equation}
with
\begin{equation}
\mathfrak{a} = 1+\hat\ell+2i \tilde m_L,\quad\mathfrak{b} = 1+\hat\ell+2i\tilde m_R,\quad\mathfrak{c}=\ell+2.
\label{eq:parhstandard}
\end{equation}
These satisfy
\begin{equation}
\mathfrak{a}+\mathfrak{b}-\mathfrak{c}=2i\left(\tilde m_L+\tilde m_R\right).
\end{equation}
We summarize some of the salient features of the hypergeometric equation and its solutions in \cref{app:2F1}.

Let us assume that $\tilde m_L$ and $\tilde m_R$ are both non-vanishing.\footnote{If $a=\pm b$ then one of these vanishes, and in the Schwarzschild--Tangherlini case ($a=b=0$) both vanish. In either of these cases, a different basis of hypergeometric solutions needs to be selected, as discussed in detail in \cref{app:2F1}. However it turns out that the solution which is regular at the horizon in each of these cases can be obtained as a limit of the general solution.} Then $\frc$ is an integer but none of $\fra$, $\frb$, and $\fra+\frb-\frc$ are, and we can choose a basis of solutions to be \cite{Bateman:100233}
\begin{align}
u_1(z) &= F(\fra,\frb;\frc;z),\\
u_2(z) &= F(\fra,\frb;1+\fra+\frb-\frc;1-z).
\end{align}
At the outer horizon $z=1$, $u_1(z)$ blows up while $u_2(z)$ is regular, so we will focus on $u_2(z)$ as the physical solution. Now we want to expand $u_2(z)$ around infinity ($z=0$). For the parameters we have chosen, the following identity holds:\footnote{See \cref{eq:integercsolutions,eq:Dabc}.}
\begin{align}
u_2(z) &= F(\fra,\frb;1+\fra+\frb-\frc;1-z) \nonumber\\
&= F(\fra,\frb;\frc;z)\ln z - \displaystyle\sum_{n=1}^{\frc-1}\frac{(\frc-1)!(n-1)!}{(\frc-n-1)!(\fra-n)_n(\frb-n)_n}(-z)^{-n} \nonumber\\
&\hphantom{{}=}+ \displaystyle\sum_{n=0}^\infty\frac{(\fra)_n(\frb)_n}{n!(\frc)_n}\left[\psi(\fra+n)+\psi(\frb+n)-\psi(1+n)-\psi(\frc+n)\right]z^n.
\end{align}
Here $(a)_k=\Gamma(a+k)/\Gamma(a)$ is the Pochhammer symbol, and $\psi(z)=\partial_z\ln\Gamma(z)$ is the digamma function.
As $z\to0$ this is dominated by the terms in the top line, in particular the log and the term in the sum with $n=\frc-1=\ell+1$:
\begin{equation}
u_2(z) \to \ln z - \frac{\ell!(\ell+1)!}{(\fra-\ell-1)_{\ell+1}(\frb-\ell-1)_{\ell+1}}(-z)^{-(\ell+1)}.
\end{equation}
The first term corresponds to the decaying $r^{-\ell-2}$ falloff in $R(r)$, and the second to the growing $r^\ell$ falloff, so the static response is given by the ratio of the first to the second coefficient:
\begin{equation}
\boxed{\lambda_\ell = 2(-1)^{\ell} \frac{(-\hat\ell+2i\tilde m_L)_{\ell+1}(-\hat\ell+2i\tilde m_R)_{\ell+1}}{\ell!(\ell+1)!}\ln\left(\frac{r_0}{r}\right).}\label{eq:LoveMP}
\end{equation}

As a check, we can compare this to the non-spinning Schwarzschild--Tangherlini metric, which is the limit $a=b=0$ (implying $\tilde m_L=\tilde m_R=0$) of the Myers--Perry solution. The induced response in this spacetime, for general $D$, are known to be zero for integer $\hl=\ell/(D-3)$ and to run logarithmically for half-integer $\hl$ \cite{Kol:2011vg,Hui:2020xxx}. In $D=5$ these are the only two options. We can see this behavior from the expression \eqref{eq:LoveMP} by inspecting the Pochhammer symbols in this limit,
\begin{equation}
(-\hat\ell)_{\ell+1} = \frac{\Gamma(\hat\ell+1)}{\Gamma(-\hat\ell)} = \begin{cases}0, & \text{$\hat\ell$ integer}, \\ (-1)^{(\ell+1)/2}\frac{\ell!!^2}{2^{\ell+1}}, & \text{$\hat\ell$ half-integer}.
\end{cases}
\end{equation}
For integer $\hat\ell$, $1/\Gamma(-\hat\ell)=0$, so that we recover the vanishing of the Love numbers. For half-integer $\hat\ell$, \cref{eq:LoveMP} agrees with eq. (4.21) of \rcite{Hui:2020xxx}.

\section{Black Ring}
\label{Sec:BR}

\subsection{Background}
In this section, we compute the Love numbers for spinning black rings in $D=5$. To this end we first review some of the properties of black rings relevant to this paper.

The black ring is a solution of vacuum Einstein's equations in five spacetime dimensions \cite{Emparan:2001wn}. In contrast with the $5D$ Myers--Perry black hole, whose horizon is topologically a 3-sphere, the black ring represents a spinning, ring-shaped object with an event horizon that is topologically a $S^1\times S^2$ . The black ring solution has several interesting properties, such as the existence of an ergosphere outside the horizon where objects can be dragged along with the  black ring rotation.

The current literature on the black ring spacetime is reviewed in \rcite{Emparan:2006mm} and, the corresponding geometry has been given in related forms in \rcite{Emparan:2001wn}. In this paper we shall work primarily with the metric in the  $(r,\theta)$ coordinates introduced in \rcite{Emparan:2006mm} and parameters $(r_0,\sigma)$ which will correspond respectively to the mass and spin parameters.
The solution is given by 
\begin{align}
\D s^2=&-\frac{\hat f}{\hat g}\left(\D t-
r_0\sinh\sigma\cosh\sigma\sqrt{\frac{R+r_0\cosh^2\sigma}{R- r_0\cosh^2\sigma}}\:
\frac{\frac{r}{R}-1}{r\hat f}\:R\:\D\psi\right)^2 \nonumber \\
&
+
\frac{\hat g}{\left(1+\frac{r\cos\theta}{R}\right)^2}
\left[
\frac{f}{\hat f}\left(1-\frac{r^2}{R^2}\right)\, R^2\D\psi^2
+
\frac{\D r^2}{(1-\frac{r^2}{R^2})f}
+ 
\frac{r^2}{g}\,\D\theta^2
+
\frac{g}{\hat g}\, r^2 \sin^2\theta\, \D\phi^2
\right],
\label{neutral}
\end{align}
where
\beq\label{ffs}
f=1-\frac{r_0}{r}\,,\qquad \hat f=1-\frac{r_0\cosh^2\sigma}{r}\,,
\eeq
and 
\beq\label{ggs}
g=1+\frac{r_0}{R} \cos\theta\,,\qquad
\hat g=1+\frac{r_0\cosh^2\sigma}{R} \cos\theta \,.
\eeq
The coordinates vary within the ranges $0\leq r\le R$, $0<\theta<\pi$ and $0\le \psi,\phi \le 2\pi$, and the dimensionless parameters within 
\beq
0< r_0 \le r_0 \cosh^2\sigma <R .
\eeq
In these coordinates, $R$ has dimensions of length, and for thin large rings it corresponds roughly to the radius of the ring $S^1$ circle. 
In order to avoid conical singularities the parameters $(r_0,\sigma)$ have to be related by $\cosh^2\sigma= 2/(1+(r_0/R)^2)$. Fixing these values leaves only two independent parameters in the solution, $R$ and $r_0$. Actually, this is to be expected based on physical principles. When you have the mass and radius of a ring, the angular momentum needs to be adjusted to achieve a balance between the tension and self-attraction of the ring with the centrifugal force. This results in only two remaining free parameters. It is easy to see that the solution has a regular outer horizon at $r=r_0$. In addition, there is an inner horizon at $r=0$ and a ring-shaped ergosurface present at $r= r_0 \cosh \sigma ^2 $.

One advantage of choosing the specific $(r,\theta)$ coordinates in the study of black rings is that the limit of a black string becomes straightforward. Consider the limit 
\beq\label{thinring}
r,\,r_0,\,r_0\cosh^2\sigma\ll R
\eeq
in which $g,\,\hat g \approx 1$, and redefine $\psi=z/R$. Then the metric in equation \eqref{neutral} becomes exactly the metric for a boosted black string, that extends along the $z$ direction with a boost parameter $\sigma$, and the horizon is located at $r=r_0$. In order to avoid conical singularities, $\psi$ must be identified with a period of $2\pi$, which results in periodic identification of the string's radius $R$: $z\sim z+2\pi R$. Consequently, the limit in equation \eqref{thinring} corresponds to the scenario where the ring's radius $R$ is significantly larger than its thickness $r_0$, with a focus on the region near the ring where $r\sim r_0$.

This precise definition clarifies the heuristic construction of a black ring as a boosted black string that has been bent into a circular shape. It also enables an approximate interpretation of $r_0, R$ and $\sigma$. The parameter $r_0$ is a measure of the radius of the $S^2$ at the horizon, and the ring's radius $R$. Hence, smaller values of $r_0 /R$ correspond to thinner rings. Additionally, $\cosh^2\sigma$ provides an estimate of the ring's rotational speed, and can be approximately identified with the local boost velocity $v=\tanh\sigma$.

\subsection{Klein--Gordon equation}

Often, when a spacetime possesses a Killing tensor, it is possible to find multiplicatively separable solutions of the KG equation. In the case of black rings, the equation seems not to be separable \cite{Durkee:2008an}. Only two specific scenarios allow for separability. The first case under consideration in this section
involves a static time-independent perturbation where $\omega$ equals zero \cite{Cardoso:2005sj}. Another scenario for the black ring, involves the infinite radius limit, where $R$ approaches infinity. This limit results in a boosted black string, which we will analyze in the following section.

To analyze the KG equation for a massless scalar for black rings it is convenient to adopt $(r,\theta)$ coordinates defined in terms of the most common employed $(x,y)$ coordinates. As we will see, the $(r,\theta)$ coordinates employed here are more convenient to show the separability of the wave equation and take straight string limit $R\rightarrow\infty$. Let us try the following ansatz:
\be
\Psi(t,r,\theta,\phi,z)=\e^{-i  \omega t +i m \phi+i \nu z} \,\left(1+\frac{r}{R}\cos\theta\right) \, \Phi(r,\theta) .
\ee
It is worth noting that neither $m$ nor $\nu$  are integers, as $(\phi,z)$ do not have periodicity $2\pi$. Below we will account for this. As found in \rcite{Chanson:2022wls}, the classical wave equation for $\Phi(r,\theta)$ becomes
\begin{multline}
\partial_r \left[ r\,(r-r_0)\left(1-\frac{r^2}{R^2}\right)\partial_r\, \Phi \right]+\frac{1}{\sin\theta}\,\partial_\theta \left[\left(1+\frac{r_0}{R}\cos\theta\right)\sin\theta \,\partial_\theta\, \Phi \right]\\
+\frac{r^2 }{(r_0-r)\left(1-\frac{r^2}{R^2}\right)(r-r_0 c_\sigma^2)}\left[\omega \, r_0 c_\sigma s_\sigma \left(1-\frac{r}{R}\right)\sqrt{\frac{1+\frac{r_0}{R}c_\sigma^2}{1-\frac{r_0}{R}c_\sigma^2}}-\nu (r-r_0 c_\sigma^2)\right]^2  \Phi 
\\
+\omega^2\frac{(1+\frac{r_0}{R}c_\sigma^2\cos\theta)^2r^3}{(1+\frac{r}{R}\cos\theta)^2(r-r_0 \,c_\sigma^2)} \, \Phi -m^2\frac{(1+\frac{r_0}{R}c_\sigma^2\cos\theta)}{(1+\frac{r_0}{R}\cos\theta)\sin^2\theta} \,\Phi+(f_r+f_{\theta})\,\Phi =0 ,
\label{KGring}
\end{multline}
where $f_r=-(2 r-r_0)\,r/R^2$, $f_{\theta}=-r_0\,\cos\theta/R$ and, for simplicity we defined $c_{X}=\cosh X$ and  $s_{X}=\sinh X$. Unfortunately, the $\omega^2$ term appears to hinder separation. To compute the static Love number however, we consider the above equation with $\omega=0$ that becomes fully separable. The remaining equations exhibits only regular singular points, which suggests that the problem can be solved locally around these points. This is the subject of the next section.

\subsection{Static responses}

In this section we will compute the static Love numbers for black rings. In the static limit  ($\omega=0$), the KG equation \eqref{KGring} reduces to a coupled system of equations:
\bea
\frac{1}{\sin\theta}\,\partial_\theta \left[\left(1+\frac{r_0}{R}\cos\theta\right)\sin\theta \,\partial_\theta\, \chi \right] -m^2\frac{(1+\frac{r_0}{R}c_\sigma^2\cos\theta)}{(1+\frac{r_0}{R}\cos\theta)\sin^2\theta} \,\chi-\frac{r_0}{R}\,\cos\theta\,\chi =-K \chi,
\label{eq:BRchi}
\eea
\bea
\partial_r \left[ r\,(r-r_0)\left(1-\frac{r^2}{R^2}\right)\partial_r\, \Phi_r \right]+ \nu^2 \frac{r ^2(r-r_0 c_\sigma^2) }{ (r_0-r)\left(1-\frac{r^2}{R^2}\right)} \Phi_r -(2 r-r_0)\,\frac{r}{R^2}\Phi_r =K\Phi_r , \label{radial}
\eea
when $\Phi(r,\theta)=\chi(\theta) \, \Phi_r(r)$. The solution to both equations involves the use of generalized Heun functions, with the separation constants $K$ serving as the eigenvalues on a sphere.
{Unlike in the case of the more familiar hypergeometric equation (see \cref{app:2F1}), the $K$ values in \eqref{eq:BRchi} and \eqref{radial} are not known in simple closed form, and can be computed only numerically or with perturbative methods.}
Alternatively, one can start by asking whether it is possible to find different near and far regions where the wave equations can be solved in terms of simple special functions, and then obtain a full solution by matching solutions in each region together along a surface of an intermediate overlap region. In the case of \eqref{eq:BRchi} and \eqref{radial}, we see that this occurs when the horizon radius $r_0$ of the black ring is smaller compared to the $S^1$ radius of ring:
\bea\label{regime}
\frac{r_0}{R}\ll 1 .
\eea
In this case, each of these equations will be solvable analytically in two regions. This is the regime we will focus on in the rest of the section.

\subsubsection*{Spheroidal equation}

Let us first focus on angular Laplacian.  In the limit \eqref{regime}, the angular equation \eqref{eq:BRchi} for the black ring reduces to
\bea
\frac{1}{\sin\theta}\,\partial_\theta \left(\sin\theta \,\partial_\theta\, \chi \right) -\left(\frac{m^2}{\sin^2\theta}-K\right) \,\chi=0\,.
\eea
The associated Legendre function which represents the solution that is regular at $\cos\theta=\pm1$ is exactly the case $m=0$. We can therefore consider with full generality the $m=0$ case. The corresponding eigenvalues are
\bea
K = \ell(\ell+1) +O(r_0)\,.
\eea
\subsubsection*{Radial equation}

We will now focus on the radial equation \eqref{radial}. To solve the differential equation we can perform an explicit matching, dividing the spacetime outside the horizon, $r_0\le r<R$, into two overlapping regions defined by the near region {($r_0\leq r \ll R$)} and the far region  {($r_0 \ll r < R$)}. 
However, for the computation of the Love numbers, the complete matching procedure is unnecessary. As we will see, solving the wave equation in the near-horizon region (with ingoing boundary conditions on the surface) and finding the transformation to the boundary will suffice to deduce the static response coefficients.

The radial wave equation \eqref{radial} in the near region, where the coordinate distance $r-r_0$ is small compared to $R$, takes the form
\bea
\Delta \partial_r \left(\Delta_0 \,\partial_r\, \Phi \right)+\left( V(r)-K \,  \Delta \right) \Phi   =0 ,
\eea
where $ \Delta \equiv  r (r- r_0)$ and $\Delta_0 \equiv r (r-r_0) (1-r_0^2/R^2)$. We can further replace 
\bea
 V(r)+K \,  \Delta \sim V(r_0)-K \,  \Delta_0\,,
 \eea
and  eq.~\eqref{radial} then becomes approximately
\begin{equation}
\partial_r\left[ r(r-r_0)\partial_r \Phi_{\text{near}}   \right] + \left[ \frac{r_0^2{\cal W}^2}{r(r-r_0)} - \ell(\ell+1)  \right]\Phi_{\text{near}} =0,
\label{blackringnearzone}
\end{equation}
 with
 \bea
 \mathcal{W}=\frac{\nu \, r_0 \sinh \sigma}{1-\frac{r_0^2}{R^2}}\,.
 \eea
The above equation contains three regular singular points and is solved by hypergeometric functions. To bring it into the standard form of the hypergeometric equation we define a new variable and perform a field redefinition:
\begin{equation}
x = \frac{r_0}{r} \, ,
\qquad
\Phi(r(x)) = (1-x)^{-i {\cal W}} x^{-\ell} u(x) \, .
\label{BR5DPhi}
\end{equation}
Then,  \cref{blackringnearzone} takes the standard hypergeometric form  \eqref{standardHyperG} with parameters
\begin{equation}
\mathfrak{a} = -\ell  \, ,
\qquad
\mathfrak{b} =  -\ell -2i {\cal W}  \, ,
\qquad
\mathfrak{c} =   -2\ell   \, .
\label{abcblacring5D}
\end{equation}
Note that $\mathfrak{a}  + \mathfrak{b} - \mathfrak{c} =  -2i {\cal W} $ is not an integer, but $\mathfrak{a}$ and $\mathfrak{c}$ are. To avoid  ambiguities in the definition of the Love numbers due to a possible uncertainty in the identification of the  response coefficients (see below), we first perform an analytic continuation $\ell\mapsto \mathbb{R}$ \cite{LeTiec:2020bos,Charalambous:2021mea}. The two linearly independent solutions are therefore given by \cref{BR5Du}.
Since now none of the numbers $\mathfrak{a}$, $\mathfrak{b}$, $\mathfrak{c}-\mathfrak{a}$, $\mathfrak{c}-\mathfrak{b}$ and $\mathfrak{c}$ is an integer, we can  use the connection formula \eqref{BR5Duexp}.
We then impose the boundary condition $\Phi\approx (1-x)^{-i {\cal W}}$ at the horizon $r_0$, which amounts to requiring that $u$ be regular at $x=1$. This fixes $C_2$ in terms of $C_1$ as \eqref{C1C2}.
Plugging back into \cref{BR5Du,BR5DPhi},
\begin{multline}
\Phi(r(x)) = C_1 (1-x)^{-i {\cal W}}  \bigg[  x^{-\ell} \hypergeom{2}{1}(-\ell ,-\ell -2i{\cal W},-2\ell;x)
\\
- \frac{\Gamma(-2\ell)\Gamma(\ell+1)\Gamma(\ell - 2i {\cal W}+1)}{\Gamma(-\ell)\Gamma(-\ell -2i {\cal W}) \Gamma(2\ell+2)}   \, x^{\ell+1} \hypergeom{2}{1}(\ell+1,\ell-2i {\cal W}+1,2\ell+2;x) 
\bigg] \, .
\label{BR5Du-2}
\end{multline}
Note that in the analytic continuation sense there is no uncertainty in how to split the external tidal field (i.e., the term $\propto x^{-\ell}$) from the response (i.e., the one $\propto x^{\ell+1}$): they both have subleading terms, resulting from expanding the hypergeometric functions in powers of $x$ in \eqref{BR5Du-2}, but for real values of $\ell$ the two series never overlap. On the other hand, if $\ell$ is integer, the source contains, at subleading order, a piece that is degenerate with the response falloff, introducing in principle an ambiguity in the definition of the Love numbers. To avoid this ambiguity, supplemented with the analytic continuation $\ell\in \mathbb{R}$, we can read off the response coefficients  from \eqref{BR5Du-2}, and only later
consider the physical value $\ell$ to be an integer. Taking the limit $x\rightarrow 0$, the response coefficients are\footnote{This seems to be consistent with eq.~(30) of \rcite{Cardoso:2005sj}.}
\begin{equation}
\lambda_{\ell \in \mathbb{R}} = - \frac{\Gamma(-2\ell)\Gamma(\ell+1)\Gamma(\ell - 2i {\cal W}+1)}{\Gamma(-\ell)\Gamma(-\ell -2i {\cal W}) \Gamma(2\ell+2)} \, .
\label{lovenumbersbr5dreal}
\end{equation}
We can now take the limit $\ell \rightarrow \mathbb{N}$. Using
\begin{equation}
\Gamma(-n+\varepsilon) = \frac{(-1)^n}{n! \, \varepsilon} + O(\varepsilon^0) \, ,
\qquad \text{for } n\in \mathbb{N} \, ,
\label{formulaGamma}
\end{equation}
we find
\begin{equation}
\boxed{\lambda^{\rm BR}_{\ell \in \mathbb{N}} = (-1)^{\ell+1} \frac{\Gamma(\ell+1)^2\Gamma(\ell - 2i {\cal W}+1)}{2 \, \Gamma(2\ell+1)  \Gamma(2\ell+2) \Gamma(-\ell -2i {\cal W})} \, .}
\label{lovenumbersbr5d}
\end{equation}
As in the four-dimensional black hole cases, the response coefficients are purely imaginary, which means that the static Love numbers vanish for the black ring (for all values of $\ell$ and $\nu$). It is instructive to take the limit ${\cal W} \rightarrow 0$ and compare our resulting expression with the induced response of a  Schwarzschild black hole. The static response coefficients \eqref{lovenumbersbr5d} vanish in this limit. This is consistent with the fact that, when ${\cal W} = 0$,  \cref{blackringnearzone} formally coincides with the equation of a massless scalar field on a four-dimensional Schwarzschild spacetime (see, e.g., eq.~(2.2) of \rcite{Hui:2021vcv} with $r_s\mapsto r_0$) and inherits  all the symmetry structure discussed in \rcite{Hui:2021vcv}.

\begin{figure}
\centering
 \includegraphics[scale=0.28]{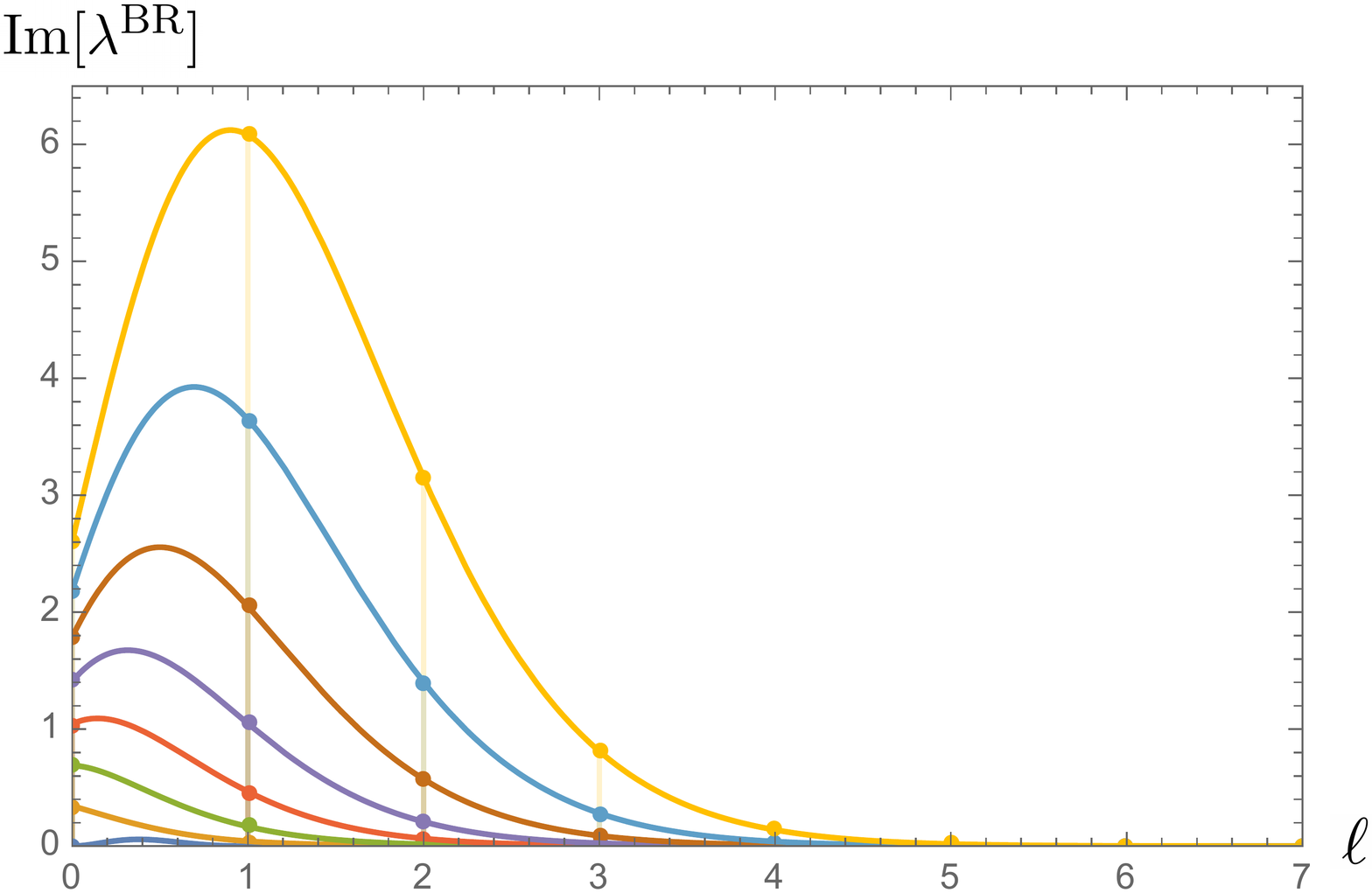}
  \caption{Visualization of the response coefficients $\lambda^{\rm BR}_{\ell}$ for black rings \eqref{lovenumbersbr5d} as a function of the multipole moments $\ell$ for various values of the boost parameter $\sigma$. The real part of the coefficients vanish, leading to vanishing static Love numbers. The non-trivial dissipation coefficients, the imaginary part of the coefficients, are represented here. As the black ring spin increases, for increasing boost parameter $\sigma$ (from {\it blue} to {\it yellow} curves), the dissipation parameters become larger. This behavior of the dissipation parameters is suppressed for increasing $\ell\rightarrow \infty$ where the coefficients vanish.}
  \label{fig1}
\end{figure}

We illustrate the behavior of the black ring configuration representing the dissipative coefficients \eqref{lovenumbersbr5d}. This is done for six different values of the angular momentum in \cref{fig1}. In each case we have plotted the dissipative part of the coefficients, $\text{Im}[\lambda^{\rm BR}_{\ell}]$, for different values of the boost parameter $\sigma$.

\section{Boosted Black Strings}
\label{Sec:BS}

The focus of this section is on boosted black string geometries, i.e., higher-dimensional stationary black string solutions carrying momentum along their length.  We will derive in particular the static response of  a test scalar field in  two distinct cases: first, we  will consider the Klein--Gordon equation on a $D$-dimensional non-rotating  boosted black string  spacetime; then, we will  focus on a boosted Kerr black string in $D=6$  and $D=5$ dimensions. 

\subsection{Non-rotating boosted black string in $D$ dimensions}
\label{subsec:bbsD}

Non-rotating boosted black string solutions in $D$ dimensions can be easily constructed by boosting the static black string metrics along the $z$ direction. The geometry of such solutions in generic $D=n+4$ dimensions is given by the following line element \cite{Emparan:2007wm,Hovdebo:2006jy}:
\begin{equation}
\begin{split}
\D s^2 =  \,  -   &  \left(1- \frac{r_0^n}{r^n} \cosh^2\sigma  \right)\D t^2 -  2\frac{r_0^n}{r^n}\cosh\sigma\sinh\sigma \, \D t \, \D z 
+ \left(1+\frac{r_0^n}{r^n} \sinh^2\sigma \right)\D z^2
\\
+ & \left( 1- \frac{r_0^n}{r^n} \right)^{-1}\D r^2 +r^2 \D \Omega_{S^{n+1}}^2 \, ,
\end{split}
\label{Dboostedstringmetric}
\end{equation}
where we assume that the $z$ direction is periodically identified \cite{Hovdebo:2006jy}.  Here $\sigma$ is the boost parameter and $\D \Omega_{S^{n+1}}^2$ the line element of the $(n+1)$-sphere $S^{n+1}$ defined recursively as $\D \Omega_{S^{n+1}}^2= \D\theta_{n+1}^2 + \sin^2\theta_{n+1} \D \Omega_{S^{n}}^2$.  This solution has an event horizon located at $r=r_0$ and an ergosurface at $r = r_0 \cosh^{2/n} \sigma$. The boost velocity is given by $v= \tanh \sigma$. The total energy and momentum of the string are, respectively,
\bea
M_{bs}=\frac{\Omega_{n+1} R}{8 G} r_0^n (n\cosh^2\sigma+1)\,,\qquad P_{bs}= \frac{\Omega_{n+1} R}{8 G} r_0^n n \cosh \sigma \sinh \sigma\,,
\eea
where $\Omega_{n+1}$ is the area of a unit $(n + 1)$-sphere. We can also define the horizon entropy
\bea
S_{bs}=\frac{\pi \, \Omega_{n+1} R}{2 G} r_0^{n+1} \cosh\sigma\,.
\eea

Let us consider a massless Klein--Gordon field $\Psi$ solving \cref{KGini} on the geometry \eqref{Dboostedstringmetric}. The symmetry structure of the metric allows us to decompose $\Psi$ in separation of variables as
\begin{equation}
\Psi (t,z,r,\theta) = \e^{-i \omega t +i \nu z }  R(r) Y_L(\theta) \, ,
\label{FTbbsn}
\end{equation}
where we Fourier transformed in time and in the coordinate $z$. $Y_L(\theta)$ are the hyperspherical harmonics with $\theta=\{\theta_1, \cdots, \theta_{n+1}\}$ the coordinates on $S^{n+1}$.\footnote{The functions  $Y_L(\theta)$ provide a representation of the rotation group SO$(n+1)$. The  dimension $N_L$ of the representation is given by $N_L = {{n+L}\choose{L}} - {{n+L-2}\choose{L-2}} = \frac{(L+n-2)!(2L+n-1)}{(n-1)!L!}$, as it can be easily found by noting that a symmetric $L$-index tensor in $(n+1)$-dimensions has ${n+L}\choose L$ independent components and that the tracelessness condition imposes ${n+L-2}\choose L-2$ conditions \cite{Hui:2020xxx}.}
Using the following expressions for the Christoffel symbols,
\begin{equation}
\Gamma^{i}_{ab} = \Gamma^a_{bi}=0 \, , \qquad 
\Gamma^{a}_{ij} = -\frac{1}{2}g^{ab}\partial_b g_{ij} \,,
\end{equation}
where the indices $i, j, \ldots$ and $a,b,\ldots$ run over the coordinates $\{t,z,r \}$ and $\{\theta_1, \cdots, \theta_{n+1}\}$ respectively, we can write the  Klein--Gordon equation \eqref{KGini}  as
\begin{equation}
 \square \Psi  = \nabla_a\nabla^a \Psi + \frac{1}{r^2} \nabla^2_{S^{n+1}}\Psi + \frac{1}{2}g^{ij }g^{ab}\partial_b g_{ij}\partial_a \Psi  =0 \, ,
\end{equation}
where $\nabla^2_{S^{n+1}}$ is the spherical Laplacian on the $S^{n+1}$ sphere.
Using that $Y_L(\theta)$ are eigenfunctions of $\nabla^2_{S^{n+1}}$ with eigenvalues
\begin{equation}
\nabla^2_{S^{n+1}} Y_L(\theta) =  -L(L+n)Y_L(\theta) \, ,
\end{equation}
the equation for the radial field component $R(r)$ can be cast in the form
\begin{equation}
r\left(r^n-r_0^n\right)  \partial_r \left[r \left(r^n-r_0^n\right) R '(r)\right]
- \left[L (L+n)r^n \left(r^n-r_0^n\right) + \nu^2 r^{n+2} \left(  r^n- r_0^n \cosh ^2\sigma  \right) \right] R(r) =0,
\label{bbsnnR}
\end{equation}
in agreement with, e.g., \rcite{Dias:2006zv} in the non-rotating limit. The equation \eqref{bbsnnR} does not admit a simple closed-form solution. However, we can introduce a near-zone approximation, defined by $r_0\leq r \ll\vert 1/\nu\vert $, where \eqref{bbsnnR} is exactly solvable in terms of hypergeometric functions, and define the induced response at values of $r$ in the range $r_0 \ll r \ll\vert 1/\nu\vert$.
In practice, we will replace $r\mapsto r_0$ in the potential as follows in such a way to preserve the form of the singularity at the horizon:
\begin{equation}
r\left(r^n-r_0^n\right)  \partial_r \left[r \left(r^n-r_0^n\right) R_{\text{near}} '(r)\right]
- \left[L (L+n)r^n \left(r^n-r_0^n\right) -  r_0^{2n}  {\cal W}^2 \right] R_{\text{near}}(r) =0 \, ,
\label{nearzoneeqn}
\end{equation}
where we defined
\begin{equation}
{\cal W} \equiv \frac{\nu r_0}{n}   \sinh\sigma  \, .
\end{equation}
We should stress that there is no unique way of defining the near-zone approximation \eqref{nearzoneeqn}.\footnote{In the context of Schwarzschild or Kerr black holes in $D=4$, see e.g.~\cite{1973JETP...37...28S,1974JETP...38....1S,Teukolsky:1973ha,Maldacena:1997ih, Lowe:2011aa, Charalambous:2021kcz,Hui:2022vbh} for some examples of near-zone approximations.} One can in fact define different schemes, corresponding to different truncations of the differential equation,  that become all exact in the limit $r\rightarrow r_0$ but  differ in the region $r>r_0$.
In this sense,  the response coefficients will be strictly speaking exact only in the limit ${\cal W}\rightarrow0$.

After the following change of coordinate and field redefinition, 
\begin{equation}
\xx = \left( \frac{r_0}{r} \right)^n \, ,
\qquad
R_{\text{near}}(r(\xx)) = (1-\xx)^{-i {\cal W}} \xx^{-\hat L} u(\xx) \, ,
\label{BR5DPhi-n}
\end{equation}
where we defined
\begin{equation}
\hat{L} \equiv \frac{L}{n} \, ,
\end{equation}
the near-zone equation \eqref{nearzoneeqn} takes the standard hypergeometric form \eqref{standardHyperG} with parameters
\begin{equation}
\mathfrak{a} = -\hat L  \, ,
\qquad
\mathfrak{b} =  -\hat L -2i {\cal W}  \, ,
\qquad
\mathfrak{c} =   -2\hat L   \, .
\label{abcblacring-n}
\end{equation}
In particular, assuming that $\hat L$ is neither integer nor semi-integer, a basis of two linearly independent solutions to the near-zone equation \eqref{nearzoneeqn} for $u(x)$ can be read off from \cref{BR5Du}.
Then, imposing the correct `infalling' boundary condition at the event horizon $r=r_0$, i.e.~(see, e.g., Refs.~\cite{Cardoso:2005sj,Dias:2006zv}),
\begin{equation}
R_{\text{near}}(r) \sim (r-r_0)^{-i {\cal W}} \, ,
\qquad
\text{as } r\rightarrow r_0\, ,
\end{equation}
is equivalent to requiring that $u(\xx)$ is finite at the singular point $\xx=1$. Such solution is written explicitly in \cref{BR5Duexp-2}. The response coefficients $\lambda_L$, defined for each $L$ as the ratio between the coefficients of the two falloffs in $R_{\text{near}}$,
\begin{equation}
R_{\text{near}}(r) \propto \left( r^L + \lambda_L \frac{r_0^{2L +n}}{r^{L+n}} \right) \, ,
\end{equation}
in the intermediate region $r_0 \ll r \ll \vert 1/\nu\vert $, across the near zone and the far zone, are thus
\begin{equation}
\boxed{\lambda_{ L} = - \frac{\Gamma(-2\hat L)\Gamma(\hat L+1)\Gamma(\hat L - 2i {\cal W}+1)}{\Gamma(-\hat L)\Gamma(-\hat L -2i {\cal W}) \Gamma(2\hat L+2)} \, ,}
\label{lovenumbersbbsn}
\end{equation}
for real $\hat L=L/n$.

Two comments are in order here. First, note that the static response coefficients \eqref{lovenumbersbbsn} of a scalar field on a boosted black string geometry in $D=n+4$ dimensions reproduce exactly the Love numbers of a Tangherlini black hole in $D-1$ dimensions \cite{Kol:2011vg,Hui:2020xxx}. Eq.~\eqref{lovenumbersbbsn} thus provides an independent check of the results of \rcite{Kol:2011vg,Hui:2020xxx} for the scalar field case. We will extend this check to rotating spacetimes in the sections below.
Second, taking the limit $D\rightarrow 5$ ($n\rightarrow1$) in  the final result \eqref{lovenumbersbbsn} recovers the static response coefficients of a black  ring in five dimensions,  to leading order in \eqref{regime}, computed in \cref{lovenumbersbr5d} (with the replacement $\ell\mapsto L$). Note that performing the calculation in generic $D$ allowed us to avoid possible ambiguities in the source/response splitting that may arise in degenerate cases and obtain an independent check of the result \eqref{lovenumbersbr5d}.

\subsection{Boosted Myers--Perry black string in 6 dimensions}

The result of the previous section can be easily extended to boosted Myers--Perry black strings. In this section, we will mainly focus on $6$-dimensional spacetimes and compute the static response of a scalar perturbation. 
 In particular, we will show that the calculation provides an alternative way of rederiving the response coefficients of a Myers--Perry black hole in $D=5$.
 
 The line element describing the geometry of a boosted Myers--Perry black string in generic $D=n+5$ dimensions\footnote{Note that, here and in the next section, the relation between $D$ and $n$ is different with respect to what we used in \cref{subsec:bbsD}.}---obtained by adding a flat direction $z$ to a Myers--Perry black hole (with single plane of rotation) and then applying a Lorentz boost to it with parameter $\sigma$---is \cite{Dias:2006zv}
\begin{equation}
\begin{split}
\D s^2  = & - \left( 1- \frac{\mu \, r^{1-n} \cosh^2\sigma}{\Sigma} \right)\D t^2 + \frac{\mu \, r^{1-n} \sinh(2\sigma)}{\Sigma}\D t \D z + \left( 1+ \frac{\mu \, r^{1-n} \sinh^2\sigma}{\Sigma} \right)\D z^2
\\
&	+ \frac{\rr\Sigma}{\Delta} \D r^2 + \Sigma \,  \D \theta^2
+ \frac{\rr(r^2+a^2)^2-\Delta a^2\sin^2\theta }{\rr\Sigma}\sin^2\theta \, \D \phi^2 
\\
&
- \frac{2\mu \, r^{1-n} \cosh \sigma}{\Sigma } a \sin^2\theta \, \D t \D \phi
 - \frac{2\mu \,  r^{1-n} \sinh\sigma}{\Sigma - \mu \, r^{1-n}} a \sin^2\theta \, \D z \D \phi +  r^2 \cos^2\theta \, \D \Omega_{S^n}^2\, ,
\end{split}
\label{boostedMPD}
\end{equation}
where $\D\Omega_{S^n}^2$ again describes the line element of a unit $n$-sphere, and
\begin{equation}
\Delta=\rr( r^2+a^2-\mu\, r^{1-n}) \, ,
\qquad
\Sigma= r^2 + a^2 \cos^2\theta\, .
\end{equation}
The Klein--Gordon equation on the geometry \eqref{boostedMPD} admits separation of variables. We shall thus decompose the (static) scalar field $\Psi$ as
\begin{equation}
\Psi = \e^{i m \phi +i \nu z }  R(r) S_{\ell}^m(\theta)Y_L(\Omega) \, ,
\end{equation}
where $S_{\ell}^m(\theta)$ are 2-dimensional spheroidal harmonics, while $Y_L$ are hyperspherical harmonics.
After straightforward manipulations, the (static) radial equation for $R(r) $ takes the form~\cite{Dias:2006zv}
\begin{equation}
\frac{\Delta}{r^{n {+2}}}\partial_{n}\left( r^{n {-2}}\Delta\partial_r R \right) + V  R=0 \, ,
\label{bMPseq}
\end{equation}
with potential
\begin{equation}
\begin{split}
V = - \frac{\Delta}{\rr} & \left[ \nu^2 r^2 
+ A_{\ell m} + L(L+n-1)\frac{a^2}{r^2} \right] + \bigg[m^2 a^2\cosh^2\sigma 
\\
&\quad  +  \mu  \, r^{1-n} (r^2+a^2)\nu^2 \sinh^2\sigma
-m^2a^2\sinh^2\sigma- 2\nu m a \mu \, r^{1-n}\sinh\sigma
\bigg]\, ,
\end{split}
\label{bMPbsV6}
\end{equation}
where $A_{\ell m} $ are the separation constants in the spheroidal harmonic equation, $A_{\ell m}=\ell(\ell+n+1)+O(a^2\nu^2)$, while $L(L+n-1)$  are the eigenvalues of the hyperspherical harmonics on the $n$-sphere. Following the logic of the previous section, we define the near zone in the range $r_+\leq r\ll \vert 1/\nu\vert$ as
\begin{equation}
\begin{split}
\frac{\Delta }{ {r^{3}}} \partial_r \left(r^{{-1}} \, \Delta   \partial_r R \right)
+ \bigg[
& - \frac{ \Delta}{\rr}  \left(\nu ^2 r_+^2 +A_{\ell m} + \frac{a^2 L^2}{r^2}\right)
+ m^2a^2  \cosh^2 \sigma 
\\
& + \mu \left(r_+^2+a^2\right)\nu^2  \sinh ^2\sigma  
- m^2a^2  \sinh ^2\sigma - 2  \nu  m a \mu  \sinh \sigma 
\bigg]R =0,
\end{split}
\label{nearboostedMP}
\end{equation}
where we set $n=1$ and where $r_+\equiv \sqrt{\mu-a^2}$ is the event horizon.
After the field redefinition and change of variable
\begin{equation}
R(r(\xx)) \equiv r^{-\frac{3}{2}} \xx^{\frac{2\ell+1}{4}+\alpha}(1-\xx)^\beta u(\xx) \, ,
\qquad\qquad
\xx\equiv \frac{r_+^2}{r^2} \, ,
\end{equation}
where
\begin{equation}
\alpha \equiv \frac{1}{2} \left(\sqrt{\nu ^2 r_+^2 +A_{\ell m}+1}-\ell-1\right) \, ,
\qquad
\beta\equiv \frac{i (a m-\nu  \mu  \sinh \sigma )}{2 r_+} \, ,
\end{equation}
the radial equation takes the standard hypergeometric form \eqref{standardHyperG}. The $\mathfrak{c}$ parameter is given by
\begin{equation}
\mathfrak{c} =  1+  \sqrt{\nu^2 r_+^2 + A_{\ell m}+1}  \, ,
\label{c6D-boostedMP}
\end{equation}
while the expressions for $\mathfrak{a}$ and $\mathfrak{b}$ are more cumbersome and we do not report  explicitly them here. We however just point out that they are non-integer---likewise $\mathfrak{c}$ in \eqref{c6D-boostedMP}---and therefore the equation belongs to the non-degenerate case with independent solutions given by \eqref{BR5Du}. To compute the response coefficients we will work perturbatively in $\nu a$, as we will eventually take the limit $\nu\rightarrow0$. We shall thus formally write:
\begin{equation}
A_{\ell m} = \ell(\ell+2) + A^{(2)}_{\ell m} \nu^2 + {\cal O}(\nu^4)\, .
\label{lambdaexp}
\end{equation}
The coefficients $A^{(2)}_{\ell m}$ can in principle  be computed, e.g., in perturbation theory or numerically, however, as we shall see below, we will not need their explicit expressions.
Since the equation is non-degenerate for $\nu\neq0$, we can take \eqref{BR5Du} with \eqref{C1C2}, which corresponds to the solution that is regular at the horizon $r=r_+$ ($\xx=1$), and expand  at large distances $r\rightarrow\infty$ ($\xx\rightarrow0$) at the first nontrivial order in $\nu$:
\begin{equation}
u(\xx\approx0) \propto  \left[ 1
- \frac{\Gamma(\mathfrak{a})\Gamma(\mathfrak{b})\Gamma(2-\mathfrak{c})}{\Gamma(\mathfrak{c})\Gamma(\mathfrak{a}-\mathfrak{c}+1)\Gamma(\mathfrak{b}-\mathfrak{c}+1)} \left( \frac{r_+}{r}\right)^{2\ell +2 + \nu^2\frac{A_{\ell m}^{(2)}+r_+^2}{ \ell +1}}  \,  \right] \, .
\label{BR6DuMP}
\end{equation}
Note that the response coefficients in \eqref{BR6DuMP} are formally divergent when $\nu=0$ because the argument of $\Gamma(2-\mathfrak{c})$ approaches a negative integer.
After using the formula \eqref{formulaGamma} 
and regularizing the result by subtracting the $1/\varepsilon$ pole term  \cite{Kol:2011vg, Hui:2020xxx}, we  find the following expression for the logarithmic dependence of the response coefficients (in units of $r_+$):\footnote{Here we are  keeping only the coefficient of the logarithmic term in the ratio between the two falloffs in \cref{BR6DuMP}. This is because only this term is unambiguous. In \cref{Love6D-MPb}, $r$ should be  thought of as the distance at which the response of the system is measured, with the logarithmic dependence being an example of classical renormalization group running~\cite{Kol:2011vg,Hui:2020xxx}. In this sense, the length scale $r_0$ plays the role of a renormalization scale to be fixed by experiments.}
\begin{equation}
\boxed{\lambda_\ell  =  \left[ (-1)^{\ell} \frac{2\Gamma(\mathfrak{a}) \Gamma(\mathfrak{b})}{\ell! \, \Gamma(\ell+2)   \Gamma(\mathfrak{a}-\ell-1) \Gamma(\mathfrak{b}-\ell-1)  }  \ln\left( \frac{r_0}{r} \right) \right]_{\nu=0} \, ,}
\label{Love6D-MPb}
\end{equation}
where the parameters  are computed at $\nu=0$.
Note that, at $\nu=0$, $\mathfrak{a}$, $\mathfrak{b}$ and $\mathfrak{c}$ coincide with \eqref{eq:parhstandard}, if in \eqref{eq:parhstandard} one sets to zero one of the two spins, e.g.~$b=0$, and identifies $m_\phi\mapsto m$ and $m_\psi\mapsto L$. In other words,  $\lambda_\ell $ in  \eqref{Love6D-MPb}  reproduces exactly the scalar response coefficients of  five-dimensional single-spin Myers--Perry black holes from \cref{eq:LoveMP}, providing  a nontrivial consistency  check of our results.

\subsection{Boosted Myers--Perry black string in 5 dimensions}

Following  the same logic of the previous section, we  now compute the static response coefficients of  a boosted Myers--Perry black string in $D=5$ dimensions (i.e., $n=0$). Taking the limit $\sigma, \nu\rightarrow0$ in the final result will allow us to rederive the dissipative response of a Kerr black hole in four dimensions~\cite{Pani:2015hfa,Landry:2015zfa,LeTiec:2020spy, LeTiec:2020bos,Charalambous:2021mea}.

In analogy with \eqref{nearboostedMP}, we first define the following near zone starting from the scalar equation \eqref{bMPseq}  (note that we need to set $L=0$ along with $n=0$):
\begin{equation}
\begin{split}
\frac{\Delta }{ {r^3}} \partial_r \left[r^{ {-1}} \, \Delta  \partial_r R \right]
+ \bigg[
&-\frac{\Delta}{\rr}  \left(\kappa ^2 r_+^2 +A_{lm} \right)
+ a^2 m^2 \cosh^2 \sigma + \mu r_+ \left(r_+^2+a^2\right) \nu^2 \sinh ^2\sigma 
\\
& 
- m^2a^2  \sinh ^2\sigma - 2  \nu  m a \mu r_+ \sinh \sigma 
\bigg]R =0\, ,
\end{split}
\label{nearboostedMP5d}
\end{equation}
where $r_+$ ($r_-$) denotes the outer (inner) horizon, obtained from solving $\Delta=0$ with $n=0$:
\begin{equation}
r_\pm \equiv \frac{\mu}{2} \pm \sqrt{\frac{\mu^2}{4} - a^2} \, .
\end{equation}
We shall then introduce 
\begin{equation}
R(r(\xx)) \equiv  \xx^{\frac{1}{2} \left(1-\sqrt{1+4  A_{\ell m} +4 \nu ^2 r_+^2}\right)}(1-\xx)^{\frac{i  \left(m a+\mu \nu r_+  \sinh \sigma \right)}{r_+-r_-}} u(\xx) \, ,
\qquad
\xx\equiv \frac{r_+-r_-}{r-r_-} \, .
\end{equation}
Then, the near-zone equation \eqref{nearboostedMP5d} takes the standard hypergeometric form \eqref{standardHyperG} with parameters
\begin{subequations}
\label{abc5D-kerr}
\begin{align}
\mathfrak{a} & = \frac{1}{2} \left(1-\sqrt{1+4 A_{\ell m}+4 \nu ^2 r_+^2}\right) \, ,
\\
\mathfrak{b} & =  \mathfrak{a} + \frac{2 i a m}{r_+-r_-} + \frac{2 i \mu a^2 \nu }{r_- (r_+-r_-)}\sinh\sigma  \, ,
\\
\mathfrak{c}  & =  2 \mathfrak{a}   \, .
\end{align}
\end{subequations}
Again, for $\nu\neq0$, none of the parameters above is an integer number.  This implies that a basis of two linearly independent solutions is \eqref{BR5Du} and the connection formula \eqref{BR5Duexp} holds. Note that imposing  the correct infalling boundary condition at the horizon is equivalent to requiring  that $u(\xx)$  is regular at $\xx=1$.\footnote{This can be easily seen for instance by recalling that, for $\sigma,\nu=0$, $R(r)$ must oscillate as $(r-r_+)^{\frac{i m a}{r_+-r_-}}$ as $r\rightarrow r_+$ \cite{1972ApJ...175..243P,Teukolsky:1973ha,Charalambous:2021mea}.}
This fixes the integration constants as in \eqref{C1C2}. Plugging back into the solution for $R$ yields
\begin{multline}
R(r(\xx)) = C_1 (1-\xx)^{\frac{i  \left(m a+\mu \kappa r_+  \sinh \sigma \right)}{r_+-r_-}} \bigg[  \xx^{\frac{\mathfrak{c}}{2}} \hypergeom{2}{1}(\mathfrak{a},\mathfrak{b},\mathfrak{c};\xx)
\\
- \frac{\Gamma(\mathfrak{c})\Gamma(\mathfrak{a}-\mathfrak{c}+1)\Gamma(\mathfrak{b}-\mathfrak{c}+1)}{\Gamma(\mathfrak{a})\Gamma(\mathfrak{b})\Gamma(2-\mathfrak{c})}    \, \xx^{1-\frac{\mathfrak{c}}{2}} \hypergeom{2}{1}(\mathfrak{a}-\mathfrak{c}+1,\mathfrak{b}-\mathfrak{c}+1,2-\mathfrak{c};\xx)
\bigg] \, .
\label{BR5Du-2kerr}
\end{multline}
Note that $\frac{\mathfrak{c}}{2}\approx -\ell  + {\cal O}(\nu^2)$ as $\nu\rightarrow0$. We can thus define the response coefficients as 
\begin{equation}
\boxed{\lambda_{\ell m} = - \frac{\Gamma(\mathfrak{c})\Gamma(\mathfrak{a}-\mathfrak{c}+1)\Gamma(\mathfrak{b}-\mathfrak{c}+1)}{\Gamma(\mathfrak{a})\Gamma(\mathfrak{b})\Gamma(2-\mathfrak{c})} \, ,}
\label{love5Dboosted}
\end{equation}
with $\mathfrak{a}$, $\mathfrak{b}$ and $\mathfrak{c}$ given in \cref{abc5D-kerr}. The static response of a scalar perturbation on Kerr spacetime can then be obtained by setting $\sigma,\nu\rightarrow0$ in \eqref{love5Dboosted}. The limit is smooth and we find:
\begin{equation}
\lambda_{\ell m}^{\text{Kerr}} = - \frac{\Gamma(-2\ell)\Gamma(\ell+1)\Gamma(1+\ell+\frac{2 i a m}{r_+-r_-})}{\Gamma(2\ell+2)\Gamma(-\ell)\Gamma(-\ell+\frac{2 i a m}{r_+-r_-})} \, ,
\label{loveKerr}
\end{equation}
correctly reproducing, e.g., eq.~(3.54) of \rcite{Charalambous:2021mea}.\footnote{Up to the factor $[(r_+-r_-)/r_s]^{2\ell+1}$ because of the slight  different definition of $\lambda_{\ell m}$.}

It is well known that an ambiguity takes place in  the calculation, in advanced Kerr coordinates, of the static response of a Kerr black hole in $D=4$~\cite{LeTiec:2020spy,LeTiec:2020bos,Charalambous:2021mea}. This happens because, similarly to what we discussed in \cref{Sec:BR}, in the physical case $\ell \in \mathbb{N}$ the subleading corrections in the falloff of the source have the same power exponent as the leading tidal response contribution. A possible way to address such  ambiguity in the source/response split is to perform an analytic continuation in $\ell$: the calculation of the response coefficients is performed by first assuming $\ell$ real, in which case the degeneracy between source and response falloffs does not occur, and by then taking in the final expression the limit of integer $\ell$~\cite{LeTiec:2020spy,LeTiec:2020bos,Charalambous:2021mea}. In this section, although similar in spirit, we provided an alternative derivation and check of the result  \eqref{loveKerr}. Doing the calculation  for a boosted black string in one higher dimension provides an alternative way of  breaking the degeneracy and defining the static response of Kerr black holes in $D=4$ \cite{Kol:2011vg}.


\section{Discussion}
\label{Sec:Discussion}

The tidal Love numbers for rotating higher-dimensional black holes capture intricate features of the dynamics of massless fields on black hole backgrounds. Our main results for Love numbers of higher-dimensional  rotating black holes can be summarized as follows:

\noindent
{\bf Conjecture:} {\it The static response coefficients $\lambda_{\ell m}$ for rotating black holes in higher-dimensional ($D\ge5$) spacetimes display the following relations
\bea
\lambda_{\ell m}^{\text{(D--1)-BH}}=\lambda_{\ell m}^{D\text{-BR}}  =\lambda_{\ell m}^{D\text{-BS}} \,,
\eea
among $(D-1)$-dimensional black holes (BH) (including Kerr and Myers--Perry black holes), $D$-dimensional thin black rings (BR) and $D$-dimensional black strings (BS).}

\begin{figure}
\centering
  \includegraphics[scale=0.3]{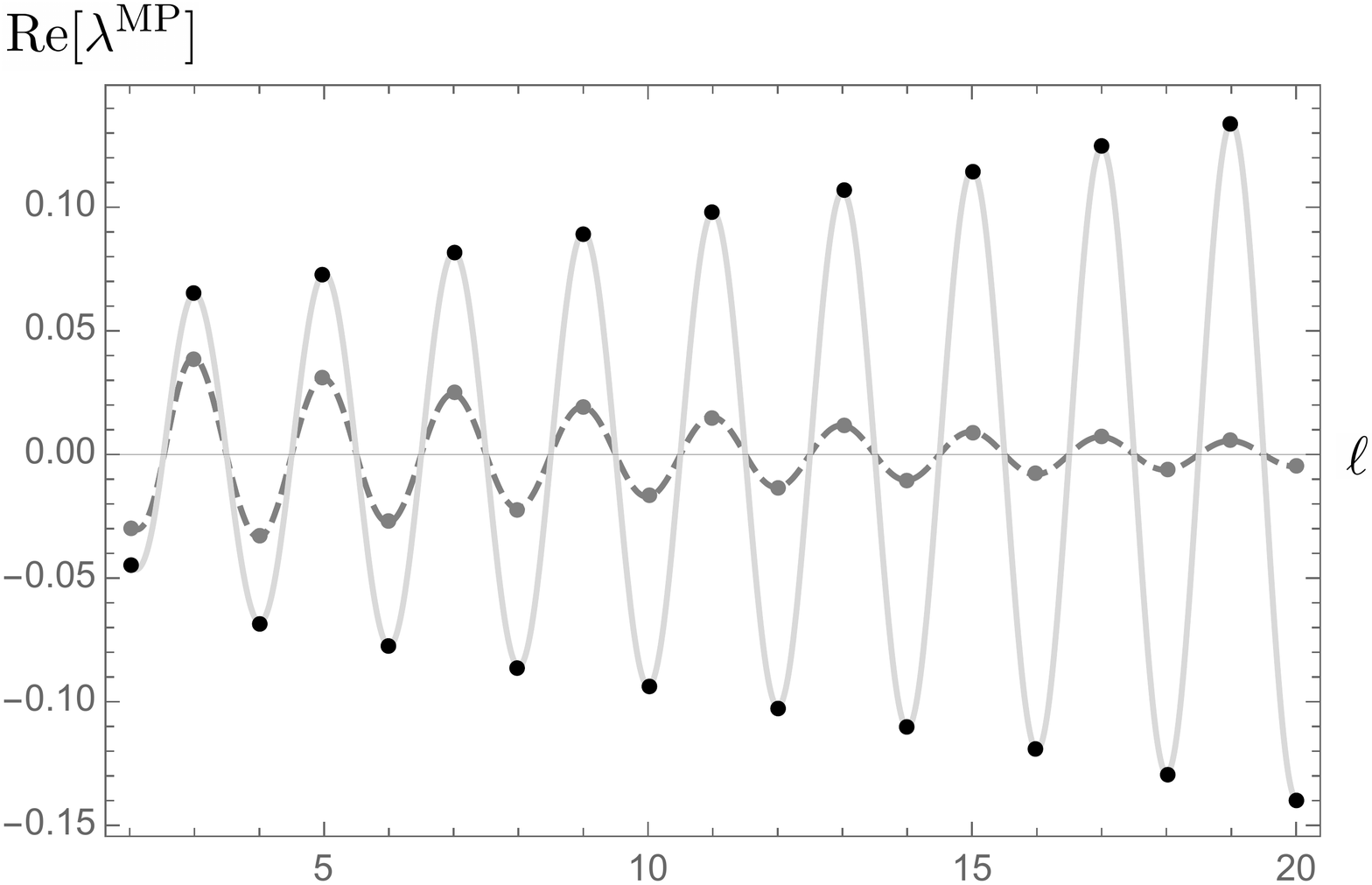}
  \caption{Visualization of the response coefficients $\lambda^{\rm MP}_{\ell m}$ for the single spinning $5D$ Myers--Perry  black holes \eqref{eq:LoveMP} as a function of the multipole moments $\ell$. The imaginary part of the coefficients vanish, leading to vanishing dissipative response coefficients. The Love numbers, defined as the real part of the $\lambda^{\rm MP}_{\ell m}$, are represented in the plot for fixed mass $M=1$ and angular momenta $J/M^2=0.26, 0.29$  (from {\it gray} to {\it black} curves), respectively below and above the critical value $(J/M^2)_\mathrm{crit} \sim 0.286$. As the multipole moments increase the Myers--Perry black holes with $J/M^2>(J/M^2)_\mathrm{crit}$ exhibit increasing values of the Love numbers. This behavior reverts for slowly rotating Myers--Perry black holes where the tidal distortion tends to zero as $\ell \rightarrow \infty$.}
  \label{fig2}
\end{figure}

From our calculation of the static Love numbers, we see that those of Myers--Perry black holes are finite, unlike their four-dimensional Kerr counterparts, which vanish. The variation of the signs of the various Love coefficients for $5D$ black holes are of paramount importance in determining the nature of the horizon and stability of the solution. 
In the presence of even/odd gravitational multipole moments, the black hole undergoes opposite distortions positive/negative. Interestingly, for, e.g., the single spinning Myers--Perry black holes ($b=0$) there seems to be a critical region around $(J/M^2)_\mathrm{crit}\sim 0.286$, where the behavior of the dissipative coefficients varies as a function of the multipole moment values $\ell$. For increasing multipole values of $\ell$, the Love numbers decrease for the slowly rotating Myers--Perry black holes with $J/M^2 < (J/M^2)_\mathrm{crit}$, while an increasing dissipative response is found for black holes with spins $J/M^2 > (J/M^2)_\mathrm{crit}$ (see \cref{fig2}). This suggests that tidal deformations for the faster spinning Myers--Perry black holes  may play an important role in elucidating the stability of these objects. Our results also complement the analysis of \rcite{Kol:2011vg,Hui:2020xxx}, which have calculated the tidal response of non-spinning Schwarzschild--Tangherlini black holes. In the limit where the spin parameters $a=b=0$ are zero the Love numbers reduce to the Schwarzschild--Tangherlini coefficients therein.

%
While the KG equation is generically not separable for black rings, exploiting the fact that for $\omega=0$ the wave equation is actually separable we were able to calculate the Love number in these backgrounds. The static response coefficients computed via a matching procedure imply the vanishing of the Love numbers for black rings, with a purely dissipative response. Importantly, we have found that the static response of  {\it thin} black rings (i.e., black rings with $r_0/R\ll 1$)  matches exactly the one of Kerr black holes. Indeed, by identifying 
\bea
\nu \, R \rightarrow m \,, \qquad  {\cal W} \rightarrow - \frac{m a}{\sqrt{r_+-r_-}} ,
\eea
the dissipative coefficients for black rings \eqref{lovenumbersbr5d} become exactly the coefficients \eqref{loveKerr} for Kerr black holes:
\begin{equation}
\lambda_{\ell m}^{\rm BR}  =\lambda_{\ell m}^{\text{Kerr}}\,,
\end{equation}
parametrized by mass $M$, spin parameter $a$, and azimuthal eigenvalue $m$. This agreement between the tidal deformation coefficients for Kerr and black rings suggests that black rings resemble much more the $4D$ black holes rather than their $5D$ counterparts, the Myers--Perry black holes. Kerr black holes have been shown to be stable \cite{giorgi2022wave} and to have a 2D CFT dual interpretation \cite{Guica:2008mu}. For black rings, stability was considered in \rcite{Hovdebo:2006jy} and a 2D CFT interpretation was also proposed \cite{Chanson:2022wls}. Our findings on the Love numbers add further evidence to the similarity between these $4D$/$5D$ black hole solutions. It will be interesting in the future to understand the connection between the tidal coefficients beyond the thin black ring regime. Finally, note that the expression for the black ring dissipation coefficients \eqref{lovenumbersbr5d} vanishes in the limit of vanishing spin parameter $\sigma \rightarrow 0$, reproducing the well-established result that the scalar response coefficients of non-spinning Schwarzschild black holes are identically zero.
 
We have also demonstrated several connections between the Love numbers for boosted black strings and $D$-dimensional black holes. Here are two points worth noting. Firstly, our analysis for the static response coefficients \eqref{lovenumbersbbsn} of a scalar field in a boosted black string geometry in $D=n+4$ dimensions perfectly reproduces the Love numbers of a Tangherlini black hole in $D-1$ dimensions \cite{Kol:2011vg,Hui:2020xxx} (see eq.~(4.17) of \rcite{Hui:2020xxx}). Equation \eqref{lovenumbersbbsn} acts as an independent verification of the outcomes of \rcite{Kol:2011vg,Hui:2020xxx} for the scalar field scenario. We also expanded this results to rotating spacetimes. Secondly, by taking the limit $D\rightarrow 5$ ($n\rightarrow1$), our expressions for the dissipation coefficients \eqref{lovenumbersbbsn} match the static response coefficients of a black ring in five dimensions \eqref{lovenumbersbr5d}. The calculation in a general number of dimensions $D$ helped us avoid potential ambiguities in the source/response splitting that arise in certain degenerate situations, and we were able obtain the coefficients without any analytic continuation in $\ell$.%

Our analysis can be extended in multiple ways. In four dimensions, a matching between the point-particle effective field theory (EFT) \cite{Goldberger:2004jt,Goldberger:2007hy} and full general relativity calculations can be defined by employing the gauge-invariant definition of Love numbers as Wilson coefficients in the EFT. Higher-dimensional gravity poses an interesting puzzle. Black holes solutions are no longer unique in vacuum, hence a complete point-particle EFT interpretation should reflect this fact. Our analysis here established connections between the different black holes tidal responses which will certainly play a role in the definitions of these coefficients as Wilson coefficients in the EFT for $5D$ gravity. In addition, we calculated response coefficients for a massless spin-0 field. The calculation of the tidal responses to spin-1 and spin-2 fields on these backgrounds has not been yet addressed.
Finally, it will be interesting to explore the (hidden) symmetries of these fields on black hole geometries in higher dimensions, and study how they constrain the tidal response of the objects~\cite{Hui:2021vcv,Hui:2022vbh}.
We leave these research directions for future work.

{\it Note added.} While this paper was being prepared, \rcite{Charalambous:2023jgq} appeared. The paper has some overlap with our work in the interpretation of Love numbers for Myers--Perry black holes in five spacetime dimensions.


\section{Acknowledgements}
We would like to thank Lam Hui, Austin Joyce, Riccardo Penco, and Malcolm Perry for useful discussions and collaboration on related topics. We thank also the Centro de Ciencias de Benasque Pedro Pascual for the hospitality where some of the research was carried out. The work of MJR is partially supported through the NSF grant PHY-2012036, RYC-2016-21159, CEX2020-001007-S and PGC2018-095976-B-C21, funded by MCIN/AEI/10.13039/501100011033. LS is supported by the Centre National de la Recherche Scientifique (CNRS).
ARS's research was partially supported by funds from the Natural Sciences and Engineering Research Council (NSERC) of Canada. Research at the Perimeter Institute is supported in part by the Government of Canada through NSERC and by the Province of Ontario through MRI. LFT acknowledges support from USU PDRF fellowship and USU Howard L. Blood Fellowship. MJR would like also to thank the Mitchell Family Foundation for hospitality in 2023 at Cook’s Branch workshop.


\appendix

\section{Some useful relations involving hypergeometric functions}
\label{app:2F1}

The hypergeometric equation is a second-order differential equation of the Fuchsian type, possessing three regular singular points. In the standard form it is written as 
\begin{equation}
\xx(1-\xx)u''(\xx) + [\mathfrak{c}-(\mathfrak{a}+\mathfrak{b}+1)\xx]u'(\xx) - \mathfrak{a} \, \mathfrak{b} \, u(\xx)=0 \, ,
\label{standardHyperG}
\end{equation}
where $ \mathfrak{a}$, $ \mathfrak{b}$ and $ \mathfrak{c}$ are constant parameters. In this appendix, we  will provide a summary of the relevant properties of the solutions and the connection coefficients in the two main cases that we encountered in the main text: \textit{(i)} none of $ \mathfrak{a}$, $ \mathfrak{b}$, $\mathfrak{c} -\mathfrak{a}$, $\mathfrak{c} -\mathfrak{b}$,  $ \mathfrak{c}$ is an integer and the equation is non-degenerate; \textit{(ii)}  $ \mathfrak{a}$, $ \mathfrak{b}$, $\mathfrak{c} -\mathfrak{a}-\mathfrak{b}$ are non-integer, while $\mathfrak{c}$ is integer. For a more complete discussion, see \rcite{Bateman:100233}.

\paragraph{Non-degenerate hypergeometric equation.}
Let us assume that none of the numbers $ \mathfrak{a}$, $ \mathfrak{b}$, $\mathfrak{c} -\mathfrak{a}$, $\mathfrak{c} -\mathfrak{b}$,  $ \mathfrak{c}$ is an integer.
In this case, the equation is non-degenerate and the two linearly independent solutions, scaling as $\sim 1 $ and $\sim \xx^{1-\mathfrak{c}}$ near the singularity $\xx=0$, are (see, e.g., Refs.~\cite{Bateman:100233,beals_wong_2010})
\begin{equation}
u(\xx) = C_1 \hypergeom{2}{1}(\mathfrak{a},\mathfrak{b},\mathfrak{c};\xx)
+ C_2 \, \xx^{1-\mathfrak{c}} \hypergeom{2}{1}(\mathfrak{a}-\mathfrak{c}+1,\mathfrak{b}-\mathfrak{c}+1,2-\mathfrak{c};\xx) \, .
\label{BR5Du}
\end{equation}
Using hypergeometric connection formulas, it is possible to re-express the linear combination \eqref{BR5Du} in terms of the fundamental solutions  in the neighborhood of any of the other two singular points. For instance, if we are interested in  $\xx=1$,  we can write
\begin{multline}
u(\xx) = C_1 \bigg[ \frac{\Gamma(\mathfrak{c})\Gamma(\mathfrak{c}-\mathfrak{a}-\mathfrak{b})}{\Gamma(\mathfrak{c}-\mathfrak{a})\Gamma(\mathfrak{c}-\mathfrak{b})} \hypergeom{2}{1}(\mathfrak{a},\mathfrak{b},\mathfrak{a}+\mathfrak{b}-\mathfrak{c}+1;1-\xx)
\\
+ \frac{\Gamma(\mathfrak{c})\Gamma(\mathfrak{a}+\mathfrak{b}-\mathfrak{c})}{\Gamma(\mathfrak{a})\Gamma(\mathfrak{b})} (1-\xx)^{\mathfrak{c}-\mathfrak{a}-\mathfrak{b}} \hypergeom{2}{1}(\mathfrak{c}-\mathfrak{a},\mathfrak{c}-\mathfrak{b},1+\mathfrak{c}-\mathfrak{a}-\mathfrak{b};1-\xx)  \bigg]
\\
+ C_2 \, \xx^{1-\mathfrak{c}} \bigg[   \frac{\Gamma(2-\mathfrak{c})\Gamma(\mathfrak{c}-\mathfrak{a}-\mathfrak{b})}{\Gamma(1-\mathfrak{a})\Gamma(1-\mathfrak{b})} \hypergeom{2}{1}(\mathfrak{a}-\mathfrak{c}+1,\mathfrak{b}-\mathfrak{c}+1,\mathfrak{a}+\mathfrak{b}-\mathfrak{c}+1;1-\xx)
\\
+ \frac{\Gamma(2-\mathfrak{c})\Gamma(\mathfrak{a}+\mathfrak{b}-\mathfrak{c})}{\Gamma(\mathfrak{a}-\mathfrak{c}+1)\Gamma(\mathfrak{b}-\mathfrak{c}+1)} (1-\xx)^{\mathfrak{c}-\mathfrak{a}-\mathfrak{b}} \hypergeom{2}{1}(1-\mathfrak{a},1-\mathfrak{b},1+\mathfrak{c}-\mathfrak{a}-\mathfrak{b};1-\xx) 
\bigg] \, ,
\label{BR5Duexp}
\end{multline}
which holds identically. In several cases in the main text, we will require that  $u(\xx)$ is regular at $\xx=1$.  This fixes $C_2$ in terms of $C_1$ as
\begin{equation}
C_2 = -C_1 \frac{\Gamma(\mathfrak{c})\Gamma(\mathfrak{a}-\mathfrak{c}+1)\Gamma(\mathfrak{b}-\mathfrak{c}+1)}{\Gamma(\mathfrak{a})\Gamma(\mathfrak{b})\Gamma(2-\mathfrak{c})} \, .
\label{C1C2}
\end{equation}
Plugging it back into \eqref{BR5Duexp} yields
\begin{multline}
u(\xx) = C_1 \frac{\Gamma(\mathfrak{c})\Gamma(\mathfrak{c}-\mathfrak{a}-\mathfrak{b})}{\Gamma(\mathfrak{c}-\mathfrak{a})\Gamma(\mathfrak{c}-\mathfrak{b})}  \bigg[ \hypergeom{2}{1}(\mathfrak{a},\mathfrak{b},\mathfrak{a}+\mathfrak{b}-\mathfrak{c}+1;1-\xx)
\\
- \frac{\Gamma(\mathfrak{c}-\mathfrak{a})\Gamma(\mathfrak{c}-\mathfrak{b}) \Gamma(\mathfrak{a}-\mathfrak{c}+1)\Gamma(\mathfrak{b}-\mathfrak{c}+1)}{\Gamma(\mathfrak{a})\Gamma(\mathfrak{b})\Gamma(1-\mathfrak{a})\Gamma(1-\mathfrak{b})} \, \xx^{1-\mathfrak{c}}  \hypergeom{2}{1}(\mathfrak{a}-\mathfrak{c}+1,\mathfrak{b}-\mathfrak{c}+1,\mathfrak{a}+\mathfrak{b}-\mathfrak{c}+1;1-\xx)
\bigg] \, .
\label{BR5Duexp-2}
\end{multline}

\paragraph{Degenerate case: integer $\mathfrak{c}$.}
Let us assume that in the hypergeometric equation \eqref{standardHyperG} the parameters $ \mathfrak{a}$, $ \mathfrak{b}$, $\mathfrak{c} -\mathfrak{a}-\mathfrak{b}$ are non-integer, while $\mathfrak{c}$ is an integer number. In such a case, the fundamental solutions in \eqref{BR5Du} are no longer independent. 
In fact, a degeneracy occurs and a basis of linearly independent hypergeometric solutions is given by  
\begin{equation}\label{eq:integercsolutions}
u_1(\xx)= \hypergeom{2}{1}(\mathfrak{a},\mathfrak{b},\mathfrak{c};\xx) \, ,
\qquad\qquad
u_2(\xx)= \ln(\xx) \frac{\hypergeom{2}{1}(\mathfrak{a},\mathfrak{b},\mathfrak{c};\xx)}{\Gamma(\mathfrak{c})}
 + \mathsf{D}_{\mathfrak{a},\mathfrak{b},\mathfrak{c}}(\xx) \, ,
\end{equation}
where
\begin{equation}\label{eq:Dabc}
\begin{split}
\mathsf{D}_{\mathfrak{a},\mathfrak{b},\mathfrak{c}}(\xx) 
& = \sum_{k=0}^\infty \left[\psi(\mathfrak{a}+k)  + \psi(\mathfrak{b}+k) -\psi(k+1) -\psi(\mathfrak{c}+k)  \right] 
\frac{(\mathfrak{a})_k(\mathfrak{b})_k}{(\mathfrak{c}-1+k)!k!} \xx^k
\\
&\quad + \sum_{k=1}^{\mathfrak{c}-1}(-1)^{k-1}\frac{(k-1)!(\mathfrak{a})_{-k}(\mathfrak{b})_{-k}}{(\mathfrak{c}-1-k)!}\xx^{-k} \, ,
\end{split}
\end{equation}
where $( \cdot)_k $ is the Pochhammer's symbol defined by $( c)_k \equiv \Gamma(c+k)/\Gamma(c) $ and $\psi $ denotes the digamma function, $\psi(z)\equiv \Gamma'(z)/\Gamma(z)$.
Now, the solution that is regular at the singular point $\xx=1$ is $u_2(\xx)$, while $u_1(\xx)$ diverges. Expanding $u_2(x)$ around $\xx=0$ yields
\begin{equation}
u_2(\xx\rightarrow0) \approx  \frac{\ln(\xx)}{\Gamma(\mathfrak{c})}
+ (-1)^{\mathfrak{c}} (\mathfrak{c}-2)! \frac{\Gamma(\mathfrak{a}-\mathfrak{c}+1)}{\Gamma(\mathfrak{a})} \frac{\Gamma(\mathfrak{b}-\mathfrak{c}+1)}{\Gamma(\mathfrak{b})}
 \xx^{1-\mathfrak{c}} + \ldots
\label{u25D-1}
\end{equation}
For our purposes in the main text, it is useful to define from \eqref{u25D-1} the $\ln(\xx)$-dependent ratio between the coefficient of the piece that goes as $\xx^{1-\mathfrak{c}}$ and the first term:
\begin{equation}
\lambda  \equiv  (-1)^{\mathfrak{c}} \frac{\Gamma(\mathfrak{a}) \Gamma(\mathfrak{b})}{(\mathfrak{c}-2)! \, \Gamma(\mathfrak{c})   \Gamma(\mathfrak{a}-\mathfrak{c}+1) \Gamma(\mathfrak{b}-\mathfrak{c}+1)  }  \ln\left( \xx \right) \, .
\label{Love5D-1}
\end{equation}


\bibliographystyle{utphys}
\addcontentsline{toc}{section}{References}
\bibliography{biblio}

\end{document}